\documentclass[twocolumn,english,superscriptaddress]{revtex4-2}
\usepackage[T1]{fontenc}
\usepackage{amsfonts}
\usepackage{amsmath}
\usepackage[latin9]{inputenc}
\usepackage{textcomp}
\usepackage{amstext}
\usepackage{subscript}
\usepackage{graphics}
\usepackage{graphicx}
\usepackage{float}
\usepackage{color}
\usepackage{pifont}
\usepackage{pmboxdraw}
\usepackage{amssymb}
\usepackage{esint}
\usepackage{tabularx}
\usepackage[export]{adjustbox}
\usepackage{hyperref}
\usepackage{upgreek,textgreek}
\usepackage{siunitx}
\usepackage{multirow}
\usepackage{mhchem}

\makeatletter

\newcommand{\lyxmathsym}[1]{\ifmmode\begingroup\def\b@ld{bold}
  \text{\ifx\math@version\b@ld\bfseries\fi#1}\endgroup\else#1\fi}

\usepackage{hyperref}
\usepackage{upgreek,textgreek}
\hypersetup{colorlinks=true,linkcolor=blue,citecolor=blue,urlcolor=blue}
\renewcommand{\fnum@figure}{FIG.~\thefigure}
\makeatother

\usepackage{babel}
\begin{document}
\title{ Anomalous Magnetotransport in the Paramagnetic State of a Magnetic Kagome Metal EuTi$_3$Bi$_4$}
\author{Yun Shu}
\thanks {These authors contributed equally to this work.}
\affiliation{Low Temperature Physics Lab, College of Physics \& Center of Quantum
Materials and Devices, Chongqing University, Chongqing 401331, China}

\author{Xinrun Mi \textcolor{blue}{\textsuperscript{*}}}
\affiliation{Low Temperature Physics Lab, College of Physics \& Center of Quantum
Materials and Devices, Chongqing University, Chongqing 401331, China}

\author{Yuhao Wei \textcolor{blue}{\textsuperscript{*}}}
\affiliation{Low Temperature Physics Lab, College of Physics \& Center of Quantum
Materials and Devices, Chongqing University, Chongqing 401331, China}

\author{Sixue Tao \textcolor{blue}{\textsuperscript{*}}}
\affiliation{Low Temperature Physics Lab, College of Physics \& Center of Quantum
Materials and Devices, Chongqing University, Chongqing 401331, China}

\author{Aifeng Wang}
\affiliation{Low Temperature Physics Lab, College of Physics \& Center of Quantum
Materials and Devices, Chongqing University, Chongqing 401331, China}

\author{Yisheng Chai}
\affiliation{Low Temperature Physics Lab, College of Physics \& Center of Quantum
Materials and Devices, Chongqing University, Chongqing 401331, China}

\author{Dashuai Ma}
\email{madason.xin@gmail.com}
\affiliation{Low Temperature Physics Lab, College of Physics \& Center of Quantum
Materials and Devices, Chongqing University, Chongqing 401331, China}

\author{Xiaolong Yang}
\email{yangxl@cqu.edu.cn}
\affiliation{Low Temperature Physics Lab, College of Physics \& Center of Quantum
Materials and Devices, Chongqing University, Chongqing 401331, China}

\author{Mingquan He}
\email{mingquan.he@cqu.edu.cn}
\affiliation{Low Temperature Physics Lab, College of Physics \& Center of Quantum
Materials and Devices, Chongqing University, Chongqing 401331, China}
\date{\today}

\begin{abstract}
We investigate the electrical transport properties of a magnetic kagome metal EuTi$_3$Bi$_4$, which undergoes magnetic ordering below $T_\mathrm{C}=10.5$ K. Unlike typical magnets showing anomalous magnetotransport in their ordered states, EuTi$_3$Bi$_4$ exhibits unusual magnetotransport behaviors in its paramagnetic phase. Specifically, the magnetoconductivity shows a linear dependence on magnetic field at low fields below $\sim 1$ T, and the Hall conductivity undergoes a sign change below about 2 T. These behaviors resemble those observed in the charge density wave (CDW) phase of kagome metals \textit{A}\ce{V3Sb5} (\textit{A} = K, Rb, Cs). The anomalous magnetotransport in \textit{A}\ce{V3Sb5} has commonly been attributed to the possible emergence of a time-reversal symmetry breaking chiral CDW order. However, given the absence of CDW in EuTi$_3$Bi$_4$ and its manifestation exclusively in the paramagnetic state, the anomalous magnetotransport observed in EuTi$_3$Bi$_4$ is likely associated with multiband transport and/or the van Hove singularities near the Fermi level.

\end{abstract}

\maketitle
\section{Introduction}
The electrical transport properties of metals are strongly influenced by the fermiology near the Fermi energy. For instance, non-zero Berry curvature near the Fermi level can give rise to the anomalous Hall effect (AHE), which is frequently observed in various topological materials \cite{Hasan2010, Xiaoliang2011, Ren_2016}. Notably, the electronic band structure of a kagome lattice exhibits several unique features, including Dirac cones, van Hove singularities, and a flat band, making it an ideal system for studying unconventional transport properties \cite{Neupert2022}. A prime example is the kagome family $A$V$_3$Sb$_5$ ($A=$ K, Rb, Cs), which shows the coexistence of superconductivity and charge density wave (CDW) order \cite{Ortiz2019, Ortiz2020, Ortiz2021, Yin2021}. Moreover, in the CDW phase, $A$V$_3$Sb$_5$ exhibits anomalous magnetotransport behavior resembling the AHE, despite the absence of spontaneous magnetization \cite{Yang2020, Yu2021, Zheng2021_gating, Zhou2021}. The concurrence of anomalous magnetotransport and CDW order suggests that the CDW phase may possess an unconventional nature \cite{Jiang2021a, Wang2021CDW, Shumiya2021, Yang2020, Yu2021, Zheng2021_gating, Chendong2021, Zhou2021, Mielke2021, Yu2021b, Gan2021, Mi_2022, Liang2021CDW, Zhao2021a, Chen2021rotonpair, Xu2021, Li2021a, Hu2022co}. Despite intensive research, the origins of this unusual CDW behavior and its associated anomalous magnetotransport remain unresolved.

The AHE-like anomalous magnetotransport in the CDW state of $A$V$_3$Sb$_5$ has commonly been attributed to a giant AHE \cite{Yang2020, Yu2021, Zheng2021_gating, Zhou2021}. Despite the absence of long-range magnetic order, extrinsic origins such as skew scattering of spin clusters or ferromagnetic fluctuations have been proposed to explain the observed AHE in $A$V$_3$Sb$_5$ \cite{Yang2020, Yu2021, Zheng2021_gating}. Intrinsically, the emergence of a non-trivial time-reversal symmetry breaking chiral CDW order in $A$V$_3$Sb$_5$ could also be a possible origin of the unusual AHE \cite{Jiang2021a}. However, to fully resolve the giant AHE in $A$V$_3$Sb$_5$, it is necessary to subtract the local linear background from the raw data \cite{Yang2020, Yu2021, Zheng2021_gating}. Furthermore, the AHE obtained after background subtraction is found to be zero in zero field, which contradicts the existence of an intrinsic time-reversal symmetry breaking scenario.

Key features of the anomalous magnetotransport in $A$V$_3$Sb$_5$ without background subtraction include: (1) Magnetoconductivity scales linearly with magnetic field in weak fields ($< 1$ T); (2) Hall conductivity varies sublinearly with magnetic field and shows a sign reversal in small magnetic fields ($< 1$ T). Notably, in samples where the AHE-like anomalous behavior is not prominent, the magnetotransport of $A$V$_3$Sb$_5$ can be qualitatively described by multiband effects, highlighting the important role of the electronic band structure \cite{Gan2021, Mi_2022, Mi2023}. Moreover, a recent study suggests that the unique fermiology, particularly the van Hove singularities near the Fermi energy, plays a key role in understanding the anomalous magnetotransport in $A$V$_3$Sb$_5$ \cite{Koshelev}. Using an analytical model that considers the backfolded concave hexagonal Fermi surface in proximity to the van Hove singularities, the anomalous magnetotransport in $A$V$_3$Sb$_5$ can be described semi-quantitatively, further emphasizing the importance of the electronic band structure in determining transport properties \cite{Koshelev}.

\begin{figure*}[t]
\centering
\includegraphics[scale=0.55]{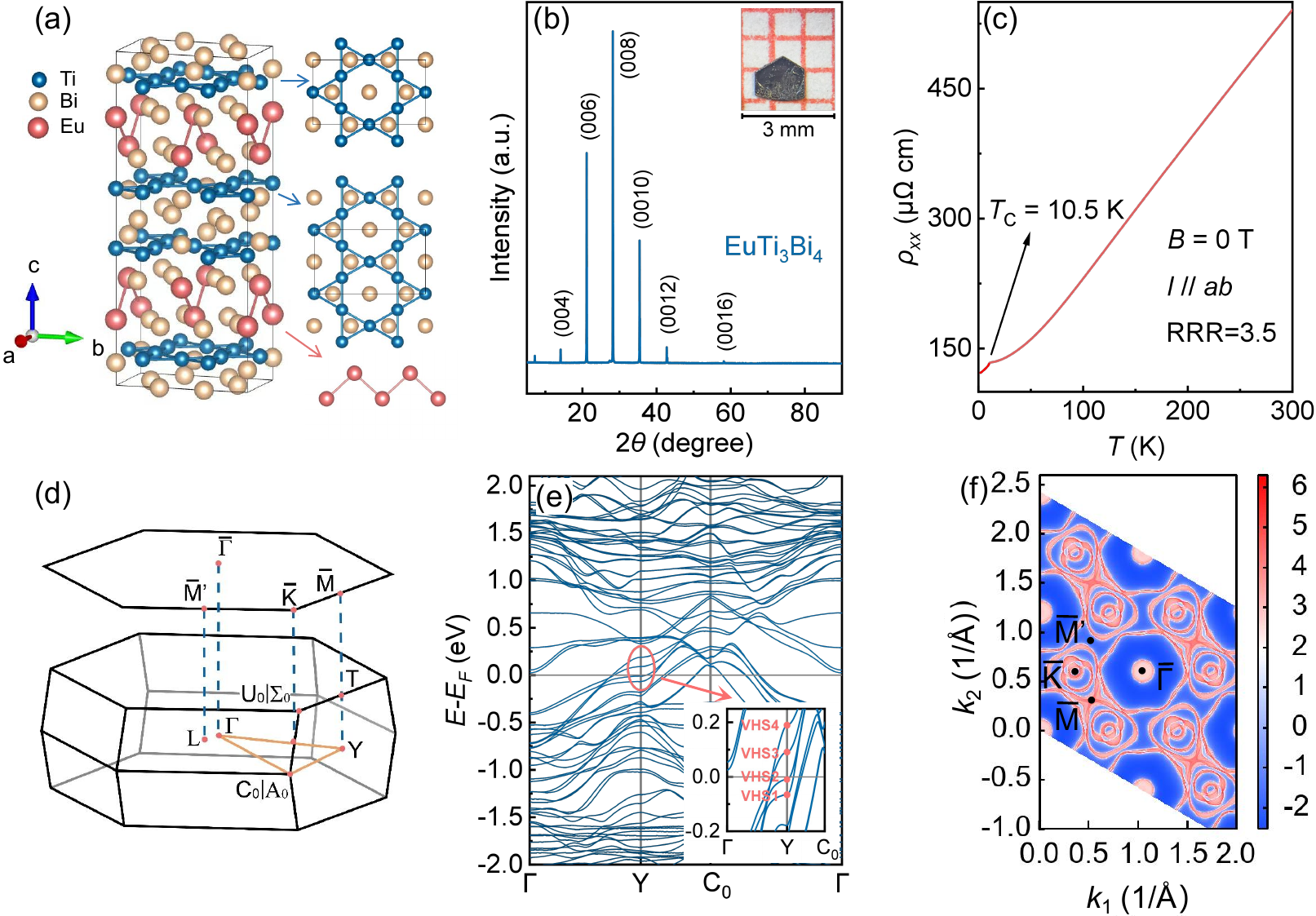}
\caption{(a) Side (left panel) and top (right panel) views of the crystal structure of \ce{EuTi3Bi4} (space group: $Fmmm$), where titanium (Ti) atoms form a distorted kagome lattice. (b) X-ray diffraction patterns of a typical \ce{EuTi3Bi4} single crystal, with clear identification of the (00$L$) peaks. The inset in (b) shows a photograph of a \ce{EuTi3Bi4} sample. (c) Temperature dependence of resistivity measured in zero magnetic field, showing a moderate residual resistivity ratio (RRR) of RRR$=\rho_{xx}$(300 K)/$\rho_{xx}$(2 K)$=3.5$, with a magnetic transition observed at $T_\mathrm{C}=10.5$ K. (d) The first Brillouin zone (bottom panel) and the two-dimensional projected surface (upper panel) of \ce{EuTi3Bi4}. (e) Calculated electronic band structure of \ce{EuTi3Bi4}, highlighting (red oval) four van Hove singularities at the Y point, with one (VHS2, -9.1 meV) located very close to the Fermi energy. (f) Projected Fermi surfaces, showing two electron-like pockets (one circular and one hexagonal) around the zone center, along with several triangular hole-like bands located at the zone corners, specifically at the $\overline{\mathrm{K}}$ point. At the $\overline{\mathrm{M}}$ point, the proximity of VHS2 to the Fermi level forms a concave pocket with sharp corners.}
\label{fig1}
\end{figure*}

In this article, we found that another kagome metal, \ce{EuTi3Bi4}, exhibits similar anomalous magnetotransport to that observed in $A$V$_3$Sb$_5$. However, unlike $A$V$_3$Sb$_5$, there is no charge density wave (CDW) in \ce{EuTi3Bi4}, and the anomalous magnetotransport is most evident in the paramagnetic state. Therefore, the anomalous magnetotransport in \ce{EuTi3Bi4} is unlikely to be associated with intrinsic time-reversal symmetry breaking. Instead, it is most likely originating from the multiband fermiology, with van Hove singularities located near the Fermi level.

\section{Experimental Method}
\ce{EuTi3Bi4} single crystals  were grown by Bi self-flux method. The purities of the materials of Eu, Ti and Bi are 99.9\%, 99.5\% and 99.99\%, respectively. The starting materials were placed in the ratio Eu: Ti: Bi = 1: 1: 20 into a alumina crucible and sealed under high vacuum in a quartz tube. The sealed ampoule was heated up to 1000$^{\circ}$C at a rate of 200/h, kept for 30 h, followed by a slow cool at 2$^{\circ}$C/h to 500$^{\circ}$C. The excess bismuth flux was removed from crystals by using a centrifuge.

Millimeter-sized shiny crystals of \ce{EuTi3Bi4} were obtained, as shown in the inset of Fig. \ref{fig1}(b). X-ray diffraction on single crystals of \ce{EuTi3Bi4} was performed using a Bruker D8 Venture diffractometer. Magnetization measurements were carried out with a Magnetic Property Measurement System (MPMS3, Quantum Design). Heat capacity measurements were conducted using the relaxation method in a Physical Property Measurement System (PPMS, Quantum Design Dynacool 9 T). Magnetoconductivity measurements were recorded using the conventional Hall bar geometry in the PPMS. The \ce{EuTi3Bi4} samples are prone to degradation when exposed to ambient air; thus, all experimental preparations were carried out under protective conditions in a glove box filled with argon.

Density functional theory (DFT) calculations \cite{Kohn1965} were performed to calculate the electronic band structure of \ce{EuTi3Bi4} using the projector augmented wave (PAW) method \cite{PhysRevB.50.17953} as implemented in the Vienna Ab Initio Simulation Package (VASP) \cite{PhysRevB.54.11169}. The Perdew-Burke-Ernzerhof (PBE) exchange-correlation functional with the generalized gradient approximation, corrected for on-site Coulomb interactions (GGA + $U$), was adopted \cite{PhysRevLett.77.3865,PhysRevLett.97.103001}. The parameter $U$ was set to 6 eV to account for the Coulomb correction from the localized 4$f$ orbitals of the Eu atoms \cite{Guo_EuTiBi}. A plane-wave cutoff energy of 450 eV was used, and the Brillouin zone (BZ) was sampled with a $7 \times 7 \times 7$ $\Gamma$-centered $k$-mesh. The local magnetic moments on the Eu atoms for the ferromagnetic (FM) state were aligned along the $z$-axis. Spin-orbit coupling (SOC) was included in all calculations. To obtain a tight-binding model for \ce{EuTi3Bi4}, we constructed a model based on maximally localized Wannier functions (MLWF), using Eu 4$f$, Ti 3$d$, and Bi 6$p$ orbitals \cite{mostofi2014updated}. The Fermi surfaces at the $k_{z}$ = 0 plane, as well as those on the two-dimensional projected surface, were calculated using the WannierTools software package \cite{wu2018wanniertools}.

\begin{figure*}
\centering  
\includegraphics[scale=0.6]{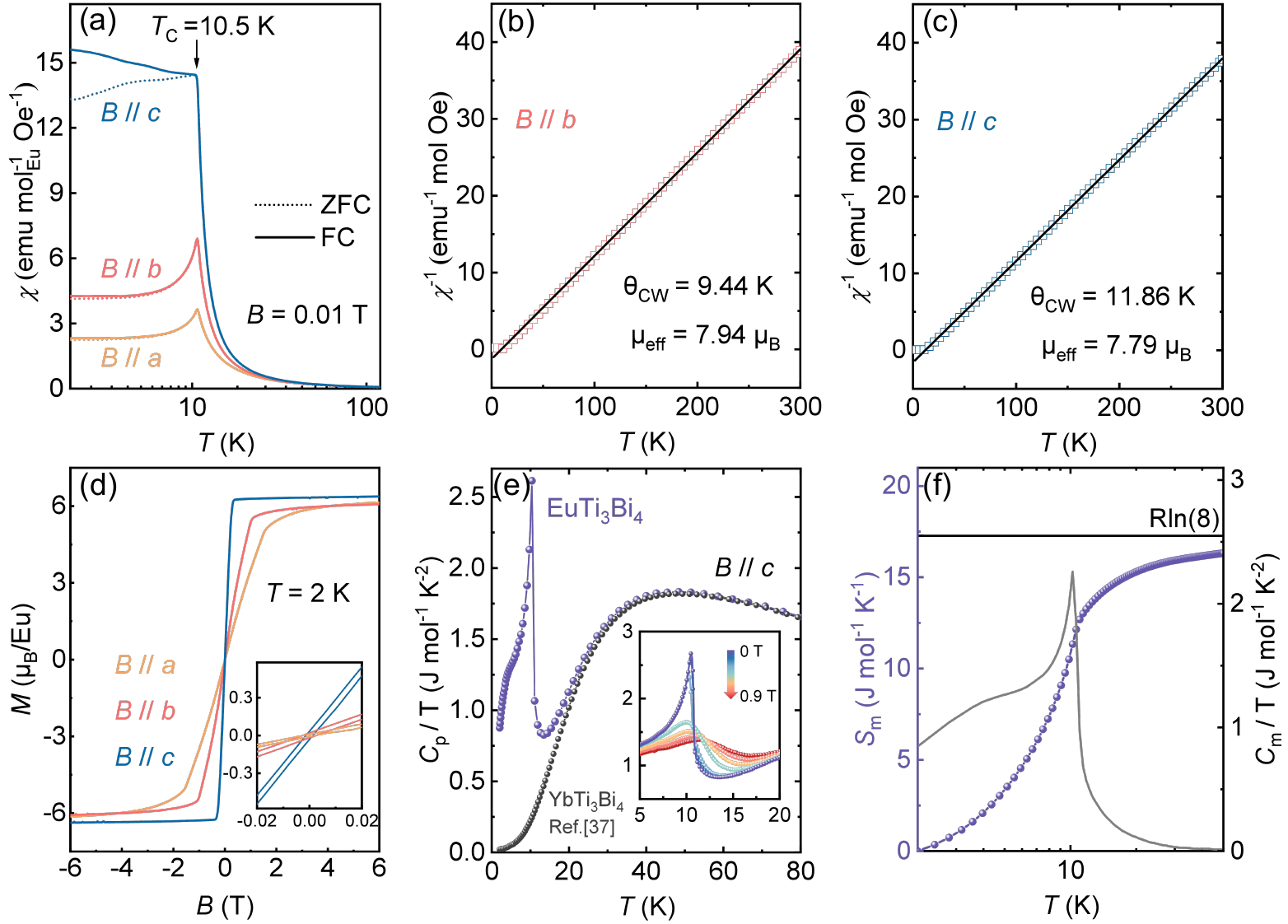}
\caption{(a) Temperature dependence of the magnetic susceptibility of \ce{EuTi3Bi4} with a magnetic field of $B=0.01$ T applied along different directions. Dashed and solid lines represent zero-field-cooled (ZFC) and field-cooled (FC) measurements, respectively. (b) and (c) Curie-Weiss analysis of the magnetic susceptibility for $B\parallel b$ and $B\parallel c$, respectively. (d) Isothermal magnetization of \ce{EuTi3Bi4} measured at 2 K with the magnetic field directed along various directions. Inset: An enlarge view near zero field, showing hysteresis in the $M-B$ curves. (e) Temperature-dependent specific heat of \ce{EuTi3Bi4}. The specific heat of \ce{YbTi3Bi4} from Ref. \cite{Ortiz_EuTiBi} is also plotted to approximate the phonon contributions. The inset in (e) shows the specific heat of \ce{EuTi3Bi4} measured under various magnetic fields. (f) Estimated magnetic specific heat ($C_\mathrm{m}$) and magnetic entropy ($S_\mathrm{m}$).}
\label{fig2}
\end{figure*}

\section{Results and Discussion}

The crystal structure of \ce{EuTi3Bi4} is shown in Fig. \ref{fig1}(a). Unlike the hexagonal $A$V$_3$Sb$_5$, \ce{EuTi3Bi4} crystallizes in an orthorhombic structure (space group No. 69, $Fmmm$) \cite{Ortiz_EuTiBi,Guo_EuTiBi,Jiang_EuTiBI}. The Ti atoms form a slightly distorted kagome lattice, with four kagome layers stacked along the $c$-axis per unit cell. The Eu atoms form quasi-1D zigzag chains running along the $a$-axis, while the Bi atoms arrange in a honeycomb pattern. The adjacent layers are loosely coupled due to the low spatial density of the zigzag chains. Thin flakes, down to a few nanometers, can be obtained by mechanical cleaving \cite{Guo_EuTiBi}. This quasi-2D structure is further supported by the weakly $k_z$-dependent band structure \cite{Guo_EuTiBi,Jiang_EuTiBI}. The weak interlayer coupling is essential for preserving the unique band structure of the Ti-kagome lattice, as shown in the calculated band structure in Fig. \ref{fig1}(e). The electronic band structure near the Fermi level is dominated by the kagome-coordinated Ti atoms. Notably, multiple van Hove singularities (VHS) appear at the Y point near the Fermi energy, as reported in earlier studies \cite{Guo_EuTiBi,Jiang_EuTiBI}. Characteristic Dirac-like bands at $\mathrm{C}_0$ and flat bands are also observed. The projected Fermi surface, shown in Fig. \ref{fig1}(f), clearly exhibits $C_2$ rotational symmetry due to the distorted Ti kagome lattice. In particular, at the $\overline{\mathrm{M}}$ point, one of the van Hove singularities (VHS2, -9.1 meV) is located very close to the Fermi level, as also evidenced by angle-resolved photoemission spectroscopy (ARPES) experiments \cite{Jiang_EuTiBI}. Meanwhile, at the $\overline{\mathrm{M}'}$ point, VHS2 is situated approximately 300 meV below the Fermi energy. The anisotropy of the Fermi surface prevents nesting between van Hove singularities, which may explain the absence of a charge density wave (CDW) instability in \ce{EuTi3Bi4}, in contrast to $A$V$_3$Sb$_5$ \cite{Jiang_EuTiBI}. Like $A$V$_3$Sb$_5$, \ce{EuTi3Bi4} is a multiband system, hosting two electron bands (one circular and one hexagonal) and several hole bands (triangular) around the center and corners of the Brillouin zone, respectively. The multiband nature plays an important role in determining the transport properties, as will be discussed later. Fig. \ref{fig1}(b) shows the X-ray diffraction pattern obtained from a typical \ce{EuTi3Bi4} single crystal. The ($00L$) peaks are clearly resolved, indicating the pure phase and [001] preferred orientation of the as-grown crystal. Temperature-dependent resistivity measurements of a \ce{EuTi3Bi4} sample reveal overall metallic behavior, with a moderate residual resistivity ratio (RRR) of 3.5, given by RRR = $\rho_{xx}$(300 K)/$\rho_{xx}$(2 K), as shown in Fig. \ref{fig2}(1). A kink in the resistivity is observed at $T_\mathrm{C} = 10.5$ K, indicating a magnetic transition.

\begin{figure*}
\centering
\includegraphics[scale=0.55]{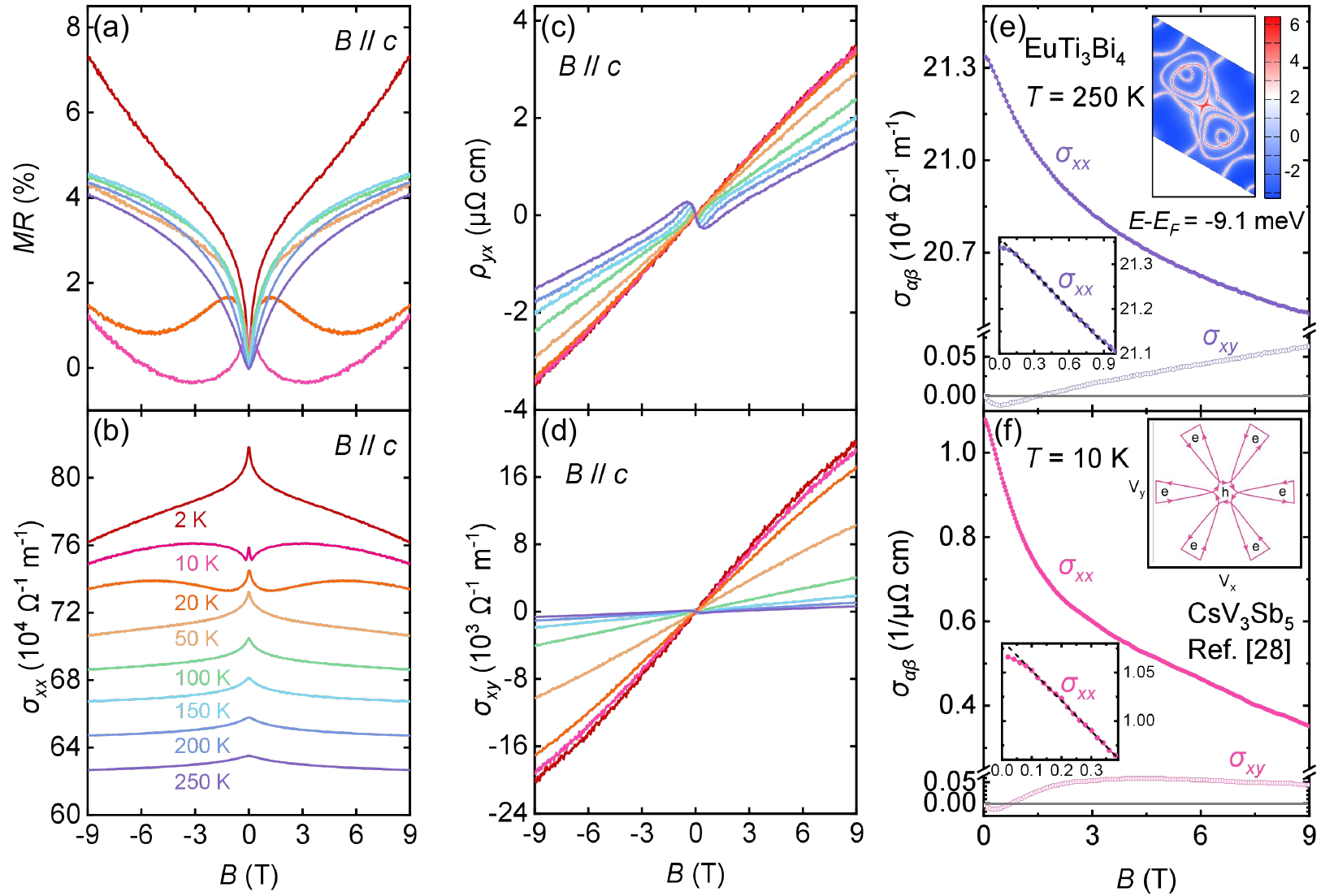}
\caption{(a) Magnetoresistance (MR) and (b) magnetoconductivity of \ce{EuTi3Bi4} measured at selected temperatures. The curves in (b) have been vertically shifted for clarity. (c) Hall resistivity and (d) Hall conductivity of \ce{EuTi3Bi4} recorded at the corresponding temperatures shown in (a) and (b). (e) and (f) Comparison of the magnetoconductivity $\sigma_{\alpha\beta}(\alpha,\beta=x,y)$ between \ce{EuTi3Bi4} and \ce{CsV3Sb5}. The data for \ce{CsV3Sb5} is adopted from Ref. \cite{Koshelev}. The inset in (e) shows the concave Fermi pocket with sharp corners near the van Hove singularity at the $\overline{\mathrm{M}}$ point. The inset in (f) shows the Fermi velocity contour of \ce{CsV3Sb5}, highlighting both hole- and electron-like contributions \cite{Koshelev}.}
\label{fig:3}
\end{figure*}

The magnetic ground state of \ce{EuTi3Bi4} remains under debate, with both ferromagnetic (FM) \cite{Ortiz_EuTiBi,Guo_EuTiBi} and antiferromagnetic (AFM) states proposed \cite{Jiang_EuTiBI}. As shown in Fig. \ref{fig2}(a), the magnetic susceptibility ($\chi$) diverges rapidly upon cooling, for both in-plane ($B \parallel ab$) and out-of-plane ($B \parallel c$) magnetic field configurations. For $B \parallel c$, $\chi$ shows only a slight change below the transition at 10.5 K. Additionally, a bifurcation appears between zero-field-cooled (ZFC) and field-cooled (FC) measurements, resembling an FM phase transition. However, for $B \parallel ab$, $\chi$ drops sharply below the transition, with no significant difference between ZFC and FC, suggesting an AFM transition. Despite this, the susceptibility for $B \parallel ab$ only decreases slightly below the transition, with a substantial net magnetization remaining. Furthermore, a finite hysteresis is observed between increasing and decreasing magnetic fields, as shown in Fig. \ref{fig2}(d). Based on these observations, we adopt the FM scenario and label the transition temperature as $T_\mathrm{C}$. Importantly, the calculated electronic band structure does not differ significantly between the FM and AFM cases, particularly near the Fermi level \cite{Guo_EuTiBi,Jiang_EuTiBI}. Above $T_\mathrm{C}$, in the paramagnetic state, the temperature-dependent magnetic susceptibility for both $B \parallel b$ and $B \parallel c$ can be well described by the Curie-Weiss law [see Figs. \ref{fig2}(b) and (c)]. The extracted Curie-Weiss temperatures ($\theta_\mathrm{CW}$) are 9.4 K and 11.8 K for $B \parallel b$ and $B \parallel c$, respectively. The positive $\theta_\mathrm{CW}$ values suggest dominant FM interactions. The estimated effective magnetic moments are $\mu_\mathrm{eff} = 7.9$ $\mu_\mathrm{B}$ for $B \parallel b$ and 7.8 $\mu_\mathrm{B}$ for $B \parallel c$, in good agreement with the theoretical value of 7.93 $\mu_\mathrm{B}$ for Eu$^{2+}$. Fig. \ref{fig2}(d) shows the isothermal magnetization measured at 2 K with magnetic field applied along different directions. For $B \parallel c$, the magnetization saturates easily in a small magnetic field of about 0.4 T, with a saturated moment of 6.3 $\mu_\mathrm{B}$. For $B \parallel a$ and $B \parallel b$, larger magnetic fields (up to 3 T) are required to fully polarize the magnetic moments. The magnetic transition is also clearly observed as a $\lambda$-shaped peak in specific heat ($C_\mathrm{p}$), shown in Fig. \ref{fig2}(e). With increasing magnetic field applied along the $c$-axis, the transition becomes broadened [see the inset in Fig. \ref{fig2}(e)]. The magnetic contribution ($C_\mathrm{m}$) to the specific heat can be estimated by subtracting the phonon background ($C_\mathrm{ph}$) from the total specific heat: $C_\mathrm{m} = C_\mathrm{p} - C_\mathrm{ph}$. Following B. R. Ortiz \textit{et al.}, we approximate the phonon contribution of \ce{EuTi3Bi4} using the specific heat of the non-magnetic, isostructural compound \ce{YbTi3Bi4} \cite{Ortiz_EuTiBi}. As shown in Fig. \ref{fig2}(e), above 50 K, the specific heat of \ce{EuTi3Bi4} matches closely with that of \ce{YbTi3Bi4}, with deviations appearing below 50 K, indicating the onset of magnetic fluctuations. The estimated $C_\mathrm{m}$, together with the magnetic entropy $S_\mathrm{m} = \int C_\mathrm{m}/T dT$, is plotted in Fig. \ref{fig2}(f). Above 50 K, $S_\mathrm{m}$ approaches the theoretical value of $R \ln(8)$ ($R$ is the ideal gas constant), suggesting that the phonon background is reasonably well approximated.

\begin{figure*}
\centering
\includegraphics[scale=0.55]{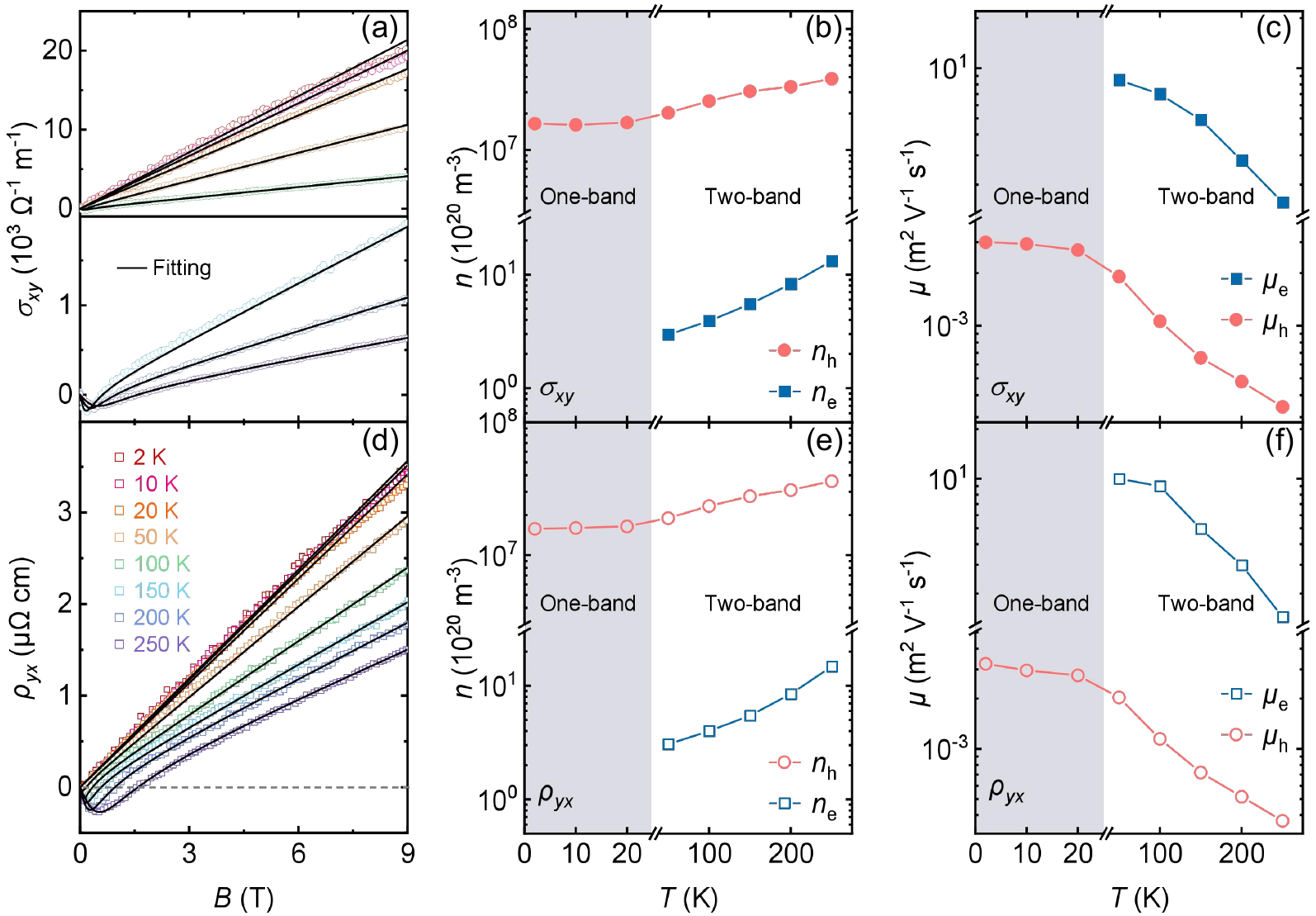}
\caption{(a, d) Two-band analysis of Hall conductivity and Hall resistivity for \ce{EuTi3Bi4}. Empty symbols represent experimental data, while solid lines are the theoretical fittings. (b, c) and (e, f) Carrier density and mobility obtained from the two-band analysis of Hall conductivity and Hall resistivity, respectively.}
\label{fig:4}
\end{figure*}

In Fig. \ref{fig:3}, we present the magnetotransport properties of \ce{EuTi3Bi4} with magnetic fields applied along the $c$-axis. Positive magnetoresistance (MR) [defined as MR $=(\rho_{xx}(B) - \rho_{xx}(0))/\rho_{xx}(0)$] is observed at most temperatures, except near $T_\mathrm{C}$. In the vicinity of $T_\mathrm{C}$, field-induced suppression of spin fluctuations may cause a weak negative magnetoresistivity in moderate magnetic fields, as commonly observed in various Eu-based magnetic materials such as \ce{EuSn2As2} \cite{Chen_2020}, \ce{EuSn2P2} \cite{xinEuSn2P2}, \ce{EuCd2As2} \cite{JunzhangEuCd2As}, \ce{EuCd2Sb2} \cite{SuEuCd2Sb2}, and \ce{EuAl2Si2} \cite{Yang2024}. Away from $T_\mathrm{C}$, the magnetoresistivity $\rho_{xx}(B)$ increases steeply in small magnetic fields (below approximately 1 T), above which it rises gradually with increasing magnetic field. These behaviors deviate from the conventional quadratic dependence of magnetoresistivity on magnetic field. At most temperatures, the magnetoconductivity $\sigma_{xx}(B)$ shows a linear dependence on magnetic field below about 1 T [see Fig. \ref{fig:3} (b) and the inset in Fig. \ref{fig:3} (e)]. Between 1 and 3 T, $\sigma_{xx}(B)$ exhibits an upward curvature, and at higher magnetic fields, $\sigma_{xx}(B)$ recovers to a quasi-linear behavior. A more interesting behavior is observed in the Hall resistivity and Hall conductivity, as shown in Figs. \ref{fig:3} (c) and (d). At first glance, the Hall data appear to be dominated by hole-like carriers in high magnetic fields. However, upon closer inspection, it is evident that above 20 K, both the Hall resistivity $\rho_{yx}(B)$ and Hall conductivity $\sigma_{xy}(B)$ vary non-monotonically with magnetic field, exhibiting a 'Z'-shaped feature. Additionally, $\rho_{yx}(B)$ and $\sigma_{xy}(B)$ change sign at a temperature-dependent magnetic field, typically below about 2 T. The 'Z'-shaped features and the sign change behavior become more pronounced with increasing temperature. By subtracting the local linear background, an AHE-like Hall response can be extracted, similar to that observed in $A$\ce{V3Sb5}. In $A$\ce{V3Sb5}, this AHE-like behavior has been attributed to a chiral charge order that breaks time-reversal symmetry \cite{Neupert2022}. However, this is unlikely the case in \ce{EuTi3Bi4} because no charge order exists in this material. Furthermore, the AHE-like Hall feature only appears well above $T_\mathrm{C}$, in the paramagnetic phase. Below 20 K, the Hall resistivity and Hall conductivity exhibit a conventional Hall effect, with $\rho_{yx}(B)$ and $\sigma_{xy}(B)$ varying monotonically with magnetic field. In the magnetic ordered state below $T_\mathrm{C}$, only the ordinary Hall effect is observed, without any signatures of AHE contributions. These findings suggest that the anomalous magnetotransport in \ce{EuTi3Bi4} is unlikely due to time-reversal symmetry breaking effects.

Figures \ref{fig:3}(e) and (f) compare the magnetoconductivity and Hall conductivity of \ce{EuTi3Bi4} and \ce{CsV3Sb5}. The data for \ce{CsV3Sb5} is adopted from the study by A. E. Koshelev \textit{et al.} \cite{Koshelev}. Notably, both the magnetoconductivity and Hall conductivity of \ce{EuTi3Bi4} resemble those of \ce{CsV3Sb5}. A. E. Koshelev \textit{et al.} suggested that the anomalous magnetotransport in \ce{CsV3Sb5} is closely associated with the van Hove singularities near the Fermi level \cite{Koshelev}. Specifically, the Fermi velocity decreases linearly as the system approaches the van Hove singularity point, giving rise to unusual magnetotransport behavior. In the pristine phase of \ce{CsV3Sb5}, a hole-like concave hexagonal pocket connects the van Hove singularities. Due to its concave geometry, the Fermi velocity contour of this hole-like pocket contains both hole-like and electron-like contributions [see the inset in Fig. \ref{fig:3}(f)]. Despite its nominal hole-like nature, the hexagonal concave pocket behaves as an effective two-band system. The reduced Fermi velocity of the hole-like part near the van Hove singularity results in dominant electron-like contributions, which manifest as negative Hall conductivity in small magnetic fields. In the CDW phase, the reconstructed Fermi surface displays similar low-field Hall conductivity to that of the pristine phase, while sharp corners in the reconstructed Fermi surface give rise to a low-field linear dependence of longitudinal magnetoconductivity \cite{Koshelev}. Given the similarities between \ce{EuTi3Bi4} and \ce{CsV3Sb5} in both longitudinal and Hall conductivity, the anomalous magnetotransport behavior in \ce{EuTi3Bi4} is also likely linked to the van Hove singularities near the Fermi energy. Although \ce{EuTi3Bi4} does not feature a closed pocket connecting the van Hove singularities due to the distorted kagome lattice, a concave pocket with sharp corners is observed near the van Hove singularity at the $\overline{M}$ point at the Fermi energy [see the inset in Fig. \ref{fig:3}(e)]. These features likely play an important role in the magnetoconductivity of \ce{EuTi3Bi4}.

Since \ce{EuTi3Bi4} is a multiband system, other Fermi pockets can also contribute to the magnetoconductivity. Moreover, the concave pocket associated with the van Hove singularity can be treated as an effective two-band model. The sign change observed around 2 T in the Hall effect is generally a consequence of multiband transport. Similar sign change effects in Hall conductivity have also been found in the kagome metal \ce{ScV6Sn6}, which can be described by multiband transport \cite{Mo_Sc,DeStefano_Sc}. In kagome metals like $A$\ce{V3Sb5} and $A$\ce{Ti3Sb5}, multiband transport also plays a crucial role in the observed transport behaviors \cite{Gan2021,Mi_2022,Mi2023,Chen_Ti3Bi5}. Here, we approximate the Hall conductivity and Hall resistivity of \ce{EuTi3Bi4} using a simple two-band model:
\begin{equation}
    \sigma_{xy}(B)=\frac{n_he\mu_h^2B}{1+\mu_h^2B^2}-\frac{n_ee\mu_e^2B}{1+\mu_e^2B^2},
    \label{eq1}
\end{equation}
\begin{equation}
   \rho_{yx}(B)=\frac{B}{e}\frac{(n_{h}\mu_{h}^{2}-n_{e}\mu_{e}^{2})+\mu_{h}^{2}\mu_{e}^{2}B^{2}(n_{h}-n_{e})}{(n_{h}\mu_{h}+n_{e}\mu_{e})^{2}+\mu_{h}^{2}\mu_{e}^{2}B^{2}(n_{h}-n_{e})^{2}},
   \label{eq2}
\end{equation}
with $n_{e(h)}$, $\mu_{e(h)}$ represent the carrier density and mobility of the corresponding electron (hole) pocket. The experimental Hall conductivity was calculated using the Hall resistivity following the relation $\sigma_{xy}=-\rho_{yx}/(\rho_{xx}^2+\rho_{yx}^2)$.  As shown in Figs. \ref{fig:4}(a) and (d), the simple two-band models described in Eqs. (\ref{eq1}) and (\ref{eq2}) can effectively describe the Hall conductivity and Hall resistivity. During the fitting, a constraint for zero-field longitudinal conductivity, $\sigma_{xx}$(0 T) $=n_h e\mu_h+n_e e\mu_e$,  was applied. The extracted carrier densities and mobilities for both types of carriers, obtained from the two-band fitting of Hall conductivity and Hall resistivity, are presented in Figs. \ref{fig:4}(b, c) and (e, f). The parameters estimated from Hall conductivity and Hall resistivity are in good agreement. Above 20 K, both electron-like and hole-like carriers contribute to the Hall effect, with low-density, high-mobility electron-like carriers dominating the low-field behavior. Below 20 K, both Hall resistivity and Hall conductivity are dominated by hole-like transport.   

\begin{figure*}
\centering
\includegraphics[scale=0.6]{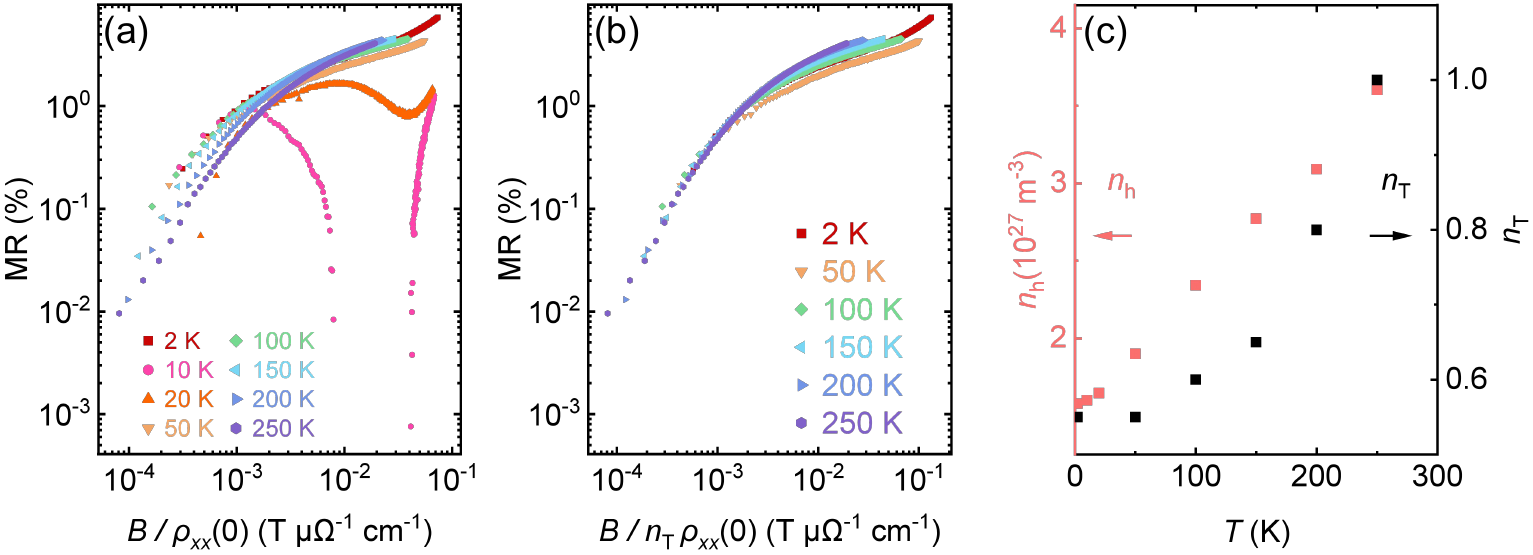}
\caption{(a) Kohler's rule scaling of MR versus $B/\rho_{xx}(0)$ for \ce{EuTi3Bi4}. (b) Extended Kohler's rule scaling with MR plotted against $B/n_\mathrm{T}\rho_{xx}(0)$. (c) Comparison of the temperature-dependent $n_\mathrm{T}$ and carrier density of the hole-like carrier, $n_\mathrm{h}$.}
\label{fig:5}
\end{figure*}

It is important to note that the low-field linear dependence of longitudinal magnetoconductivity cannot be described by the simple two-band model. This unusual behavior is likely due to sharp corners in the Fermi pockets associated with the van Hove singularity, as discussed above \cite{Koshelev}. Additionally, Kohler's rule for magnetoresistance (MR) in metals is violated in \ce{EuTi3Bi4}. Kohler's rule suggests that MR of metals follows a scaling behavior, MR$=f[B/\rho_{xx}(0)]$, where MR is a function of $B/\rho_{xx}(0)$ \cite{Kohler}. However, as shown in Fig. \ref{fig:5}(a), the scaling of MR versus $B/\rho_{xx}(0)$ measured at different temperatures does not collapse onto a single curve in \ce{EuTi3Bi4}, indicating a violation of Kohler's rule. Violations of Kohler's rule have been observed in various materials, from strongly correlated systems to topological semimetals \cite{Harris_vio, Anderson_vio, Xiang_vio, R_vio, Wu_vio, Xu_vio}. Notably, these violations can often be calibrated using an extended form of Kohler's rule: MR=$=f[B/n_\mathrm{T}\rho_{xx}(0)]$, where $n_\mathrm{T}$ represents the temperature-dependent relative variations in carrier density \cite{Xu_vio}. The temperature response of the carrier density can result from gap opening at the Fermi level during phase transitions or from thermal excitations in semiconductors and semimetals \cite{Xu_vio}. The extended Kohler's rule has been successfully applied to the MR observed in \ce{ScV6Sn6}, where the temperature-dependent $n_\mathrm{T}$ is likely a consequence of the formation of a pseudogap associated with the CDW transition \cite{DeStefano_Sc}. In Fig. \ref{fig:5}(b), we apply the extended Kohler's rule to the MR data of \ce{EuTi3Bi4}. Here, $n_\mathrm{T}$ is set to 1 at 250 K, and the values of $n_\mathrm{T}$ at other temperatures are adjusted to scale the linear part of all MR curves. The MR curves near $T_\mathrm{C}$ are excluded from this analysis, as Kohler's rule cannot account for the negative MR observed at these temperatures. The extracted temperature-dependent $n_\mathrm{T}$ is shown in Fig. \ref{fig:5}(c), alongside the temperature-dependent carrier density of the dominant hole-like carriers. The trends of $n_\mathrm{T}$ and carrier density are similar, both decreasing as temperature decreases. Since no gap opening is observed in \ce{EuTi3Bi4}, the temperature-dependent variation in carrier density is likely due to thermal excitations.

\section{Conclusions}
We have studied the electrical transport properties of the metallic kagome magnet \ce{EuTi3Bi4}. Both the longitudinal and Hall magnetoconductivity exhibit anomalous transport behaviors. Specifically, the low-field longitudinal magnetoconductivity shows a linear dependence on magnetic field, while the Hall magnetoconductivity undergoes a sign change at small magnetic fields, similar to the behaviors observed in $A$\ce{V3Sb5}. Notably, these anomalous magnetotransport properties in \ce{EuTi3Bi4} manifest in the paramagnetic state, suggesting that time-reversal symmetry breaking effects are not responsible. Instead, the multiband electronic structure and/or the presence of van Hove singularities near the Fermi level play a key role in the observed anomalous magnetotransport behavior.

\section{Acknowledgments}
 We thank A. E. Koshelev, Xianxin Wu and Yong Hu for fruitful discussions. We thank Yan Liu and Guiwen Wang at the Analytical and Testing Center of Chongqing University for their technical support. This work has been supported by National Natural Science Foundation of China (Grant No. 12104254), the Open Fund of the China Spallation Neutron Source Songshan Lake Science City (Grant No. KFKT2023B07), Fundamental Research Funds for the Central Universities, China (Grant No. 2024CDJXY022),  Chinesisch-Deutsche Mobilit\"atsprogamm of Chinesisch-Deutsche Zentrum f\"ur Wissenschaftsf\"orderung (Grant No. M-0496). X. Yang acknowledges the support by  National Natural Science Foundation of China (Grant No. 12374038).

\bibliographystyle{apsrev4-2}

\begin{thebibliography}{53}%
\makeatletter
\providecommand \@ifxundefined [1]{%
 \@ifx{#1\undefined}
}%
\providecommand \@ifnum [1]{%
 \ifnum #1\expandafter \@firstoftwo
 \else \expandafter \@secondoftwo
 \fi
}%
\providecommand \@ifx [1]{%
 \ifx #1\expandafter \@firstoftwo
 \else \expandafter \@secondoftwo
 \fi
}%
\providecommand \natexlab [1]{#1}%
\providecommand \enquote  [1]{``#1''}%
\providecommand \bibnamefont  [1]{#1}%
\providecommand \bibfnamefont [1]{#1}%
\providecommand \citenamefont [1]{#1}%
\providecommand \href@noop [0]{\@secondoftwo}%
\providecommand \href [0]{\begingroup \@sanitize@url \@href}%
\providecommand \@href[1]{\@@startlink{#1}\@@href}%
\providecommand \@@href[1]{\endgroup#1\@@endlink}%
\providecommand \@sanitize@url [0]{\catcode `\\12\catcode `\$12\catcode `\&12\catcode `\#12\catcode `\^12\catcode `\_12\catcode `\%12\relax}%
\providecommand \@@startlink[1]{}%
\providecommand \@@endlink[0]{}%
\providecommand \url  [0]{\begingroup\@sanitize@url \@url }%
\providecommand \@url [1]{\endgroup\@href {#1}{\urlprefix }}%
\providecommand \urlprefix  [0]{URL }%
\providecommand \Eprint [0]{\href }%
\providecommand \doibase [0]{https://doi.org/}%
\providecommand \selectlanguage [0]{\@gobble}%
\providecommand \bibinfo  [0]{\@secondoftwo}%
\providecommand \bibfield  [0]{\@secondoftwo}%
\providecommand \translation [1]{[#1]}%
\providecommand \BibitemOpen [0]{}%
\providecommand \bibitemStop [0]{}%
\providecommand \bibitemNoStop [0]{.\EOS\space}%
\providecommand \EOS [0]{\spacefactor3000\relax}%
\providecommand \BibitemShut  [1]{\csname bibitem#1\endcsname}%
\let\auto@bib@innerbib\@empty
\bibitem [{\citenamefont {Hasan}\ and\ \citenamefont {Kane}(2010)}]{Hasan2010}%
  \BibitemOpen
  \bibfield  {author} {\bibinfo {author} {\bibfnamefont {M.~Z.}\ \bibnamefont {Hasan}}\ and\ \bibinfo {author} {\bibfnamefont {C.~L.}\ \bibnamefont {Kane}},\ }\href {https://doi.org/10.1103/RevModPhys.82.3045} {\bibfield  {journal} {\bibinfo  {journal} {Rev. Mod. Phys.}\ }\textbf {\bibinfo {volume} {82}},\ \bibinfo {pages} {3045} (\bibinfo {year} {2010})}\BibitemShut {NoStop}%
\bibitem [{\citenamefont {Qi}\ and\ \citenamefont {Zhang}(2011)}]{Xiaoliang2011}%
  \BibitemOpen
  \bibfield  {author} {\bibinfo {author} {\bibfnamefont {X.-L.}\ \bibnamefont {Qi}}\ and\ \bibinfo {author} {\bibfnamefont {S.-C.}\ \bibnamefont {Zhang}},\ }\href {https://doi.org/10.1103/RevModPhys.83.1057} {\bibfield  {journal} {\bibinfo  {journal} {Rev. Mod. Phys.}\ }\textbf {\bibinfo {volume} {83}},\ \bibinfo {pages} {1057} (\bibinfo {year} {2011})}\BibitemShut {NoStop}%
\bibitem [{\citenamefont {Ren}\ \emph {et~al.}(2016)\citenamefont {Ren}, \citenamefont {Qiao},\ and\ \citenamefont {Niu}}]{Ren_2016}%
  \BibitemOpen
  \bibfield  {author} {\bibinfo {author} {\bibfnamefont {Y.}~\bibnamefont {Ren}}, \bibinfo {author} {\bibfnamefont {Z.}~\bibnamefont {Qiao}},\ and\ \bibinfo {author} {\bibfnamefont {Q.}~\bibnamefont {Niu}},\ }\href {https://doi.org/10.1088/0034-4885/79/6/066501} {\bibfield  {journal} {\bibinfo  {journal} {Rep. Prog. Phys.}\ }\textbf {\bibinfo {volume} {79}},\ \bibinfo {pages} {066501} (\bibinfo {year} {2016})}\BibitemShut {NoStop}%
\bibitem [{\citenamefont {Neupert}\ \emph {et~al.}(2022)\citenamefont {Neupert}, \citenamefont {Denner}, \citenamefont {Yin}, \citenamefont {Thomale},\ and\ \citenamefont {Hasan}}]{Neupert2022}%
  \BibitemOpen
  \bibfield  {author} {\bibinfo {author} {\bibfnamefont {T.}~\bibnamefont {Neupert}}, \bibinfo {author} {\bibfnamefont {M.~M.}\ \bibnamefont {Denner}}, \bibinfo {author} {\bibfnamefont {J.-X.}\ \bibnamefont {Yin}}, \bibinfo {author} {\bibfnamefont {R.}~\bibnamefont {Thomale}},\ and\ \bibinfo {author} {\bibfnamefont {M.~Z.}\ \bibnamefont {Hasan}},\ }\href {https://doi.org/10.1038/s41567-021-01404-y} {\bibfield  {journal} {\bibinfo  {journal} {Nat. Phys.}\ }\textbf {\bibinfo {volume} {18}},\ \bibinfo {pages} {137} (\bibinfo {year} {2022})}\BibitemShut {NoStop}%
\bibitem [{\citenamefont {Ortiz}\ \emph {et~al.}(2019)\citenamefont {Ortiz}, \citenamefont {Gomes}, \citenamefont {Morey}, \citenamefont {Winiarski}, \citenamefont {Bordelon}, \citenamefont {Mangum}, \citenamefont {Oswald}, \citenamefont {Rodriguez-Rivera}, \citenamefont {Neilson}, \citenamefont {Wilson}, \citenamefont {Ertekin}, \citenamefont {McQueen},\ and\ \citenamefont {Toberer}}]{Ortiz2019}%
  \BibitemOpen
  \bibfield  {author} {\bibinfo {author} {\bibfnamefont {B.~R.}\ \bibnamefont {Ortiz}}, \bibinfo {author} {\bibfnamefont {L.~C.}\ \bibnamefont {Gomes}}, \bibinfo {author} {\bibfnamefont {J.~R.}\ \bibnamefont {Morey}}, \bibinfo {author} {\bibfnamefont {M.}~\bibnamefont {Winiarski}}, \bibinfo {author} {\bibfnamefont {M.}~\bibnamefont {Bordelon}}, \bibinfo {author} {\bibfnamefont {J.~S.}\ \bibnamefont {Mangum}}, \bibinfo {author} {\bibfnamefont {I.~W.}\ \bibnamefont {Oswald}}, \bibinfo {author} {\bibfnamefont {J.~A.}\ \bibnamefont {Rodriguez-Rivera}}, \bibinfo {author} {\bibfnamefont {J.~R.}\ \bibnamefont {Neilson}}, \bibinfo {author} {\bibfnamefont {S.~D.}\ \bibnamefont {Wilson}}, \bibinfo {author} {\bibfnamefont {E.}~\bibnamefont {Ertekin}}, \bibinfo {author} {\bibfnamefont {T.~M.}\ \bibnamefont {McQueen}},\ and\ \bibinfo {author} {\bibfnamefont {E.~S.}\ \bibnamefont {Toberer}},\ }\href {https://doi.org/10.1103/PhysRevMaterials.3.094407} {\bibfield  {journal} {\bibinfo  {journal} {Phys. Rev. Mater.}\ }\textbf
  {\bibinfo {volume} {3}},\ \bibinfo {pages} {094407} (\bibinfo {year} {2019})}\BibitemShut {NoStop}%
\bibitem [{\citenamefont {Ortiz}\ \emph {et~al.}(2020)\citenamefont {Ortiz}, \citenamefont {Teicher}, \citenamefont {Hu}, \citenamefont {Zuo}, \citenamefont {Sarte}, \citenamefont {Schueller}, \citenamefont {Abeykoon}, \citenamefont {Krogstad}, \citenamefont {Rosenkranz}, \citenamefont {Osborn}, \citenamefont {Seshadri}, \citenamefont {Balents}, \citenamefont {He},\ and\ \citenamefont {Wilson}}]{Ortiz2020}%
  \BibitemOpen
  \bibfield  {author} {\bibinfo {author} {\bibfnamefont {B.~R.}\ \bibnamefont {Ortiz}}, \bibinfo {author} {\bibfnamefont {S.~M.~L.}\ \bibnamefont {Teicher}}, \bibinfo {author} {\bibfnamefont {Y.}~\bibnamefont {Hu}}, \bibinfo {author} {\bibfnamefont {J.~L.}\ \bibnamefont {Zuo}}, \bibinfo {author} {\bibfnamefont {P.~M.}\ \bibnamefont {Sarte}}, \bibinfo {author} {\bibfnamefont {E.~C.}\ \bibnamefont {Schueller}}, \bibinfo {author} {\bibfnamefont {A.~M.~M.}\ \bibnamefont {Abeykoon}}, \bibinfo {author} {\bibfnamefont {M.~J.}\ \bibnamefont {Krogstad}}, \bibinfo {author} {\bibfnamefont {S.}~\bibnamefont {Rosenkranz}}, \bibinfo {author} {\bibfnamefont {R.}~\bibnamefont {Osborn}}, \bibinfo {author} {\bibfnamefont {R.}~\bibnamefont {Seshadri}}, \bibinfo {author} {\bibfnamefont {L.}~\bibnamefont {Balents}}, \bibinfo {author} {\bibfnamefont {J.}~\bibnamefont {He}},\ and\ \bibinfo {author} {\bibfnamefont {S.~D.}\ \bibnamefont {Wilson}},\ }\href {https://doi.org/10.1103/PhysRevLett.125.247002} {\bibfield  {journal}
  {\bibinfo  {journal} {Phys. Rev. Lett.}\ }\textbf {\bibinfo {volume} {125}},\ \bibinfo {pages} {247002} (\bibinfo {year} {2020})}\BibitemShut {NoStop}%
\bibitem [{\citenamefont {Ortiz}\ \emph {et~al.}(2021)\citenamefont {Ortiz}, \citenamefont {Sarte}, \citenamefont {Kenney}, \citenamefont {Graf}, \citenamefont {Teicher}, \citenamefont {Seshadri},\ and\ \citenamefont {Wilson}}]{Ortiz2021}%
  \BibitemOpen
  \bibfield  {author} {\bibinfo {author} {\bibfnamefont {B.~R.}\ \bibnamefont {Ortiz}}, \bibinfo {author} {\bibfnamefont {P.~M.}\ \bibnamefont {Sarte}}, \bibinfo {author} {\bibfnamefont {E.~M.}\ \bibnamefont {Kenney}}, \bibinfo {author} {\bibfnamefont {M.~J.}\ \bibnamefont {Graf}}, \bibinfo {author} {\bibfnamefont {S.~M.~L.}\ \bibnamefont {Teicher}}, \bibinfo {author} {\bibfnamefont {R.}~\bibnamefont {Seshadri}},\ and\ \bibinfo {author} {\bibfnamefont {S.~D.}\ \bibnamefont {Wilson}},\ }\href {https://doi.org/10.1103/PhysRevMaterials.5.034801} {\bibfield  {journal} {\bibinfo  {journal} {Phys. Rev. Mater.}\ }\textbf {\bibinfo {volume} {5}},\ \bibinfo {pages} {034801} (\bibinfo {year} {2021})}\BibitemShut {NoStop}%
\bibitem [{\citenamefont {Yin}\ \emph {et~al.}(2021)\citenamefont {Yin}, \citenamefont {Tu}, \citenamefont {Gong}, \citenamefont {Fu}, \citenamefont {Yan},\ and\ \citenamefont {Lei}}]{Yin2021}%
  \BibitemOpen
  \bibfield  {author} {\bibinfo {author} {\bibfnamefont {Q.}~\bibnamefont {Yin}}, \bibinfo {author} {\bibfnamefont {Z.}~\bibnamefont {Tu}}, \bibinfo {author} {\bibfnamefont {C.}~\bibnamefont {Gong}}, \bibinfo {author} {\bibfnamefont {Y.}~\bibnamefont {Fu}}, \bibinfo {author} {\bibfnamefont {S.}~\bibnamefont {Yan}},\ and\ \bibinfo {author} {\bibfnamefont {H.}~\bibnamefont {Lei}},\ }\href {https://doi.org/10.1088/0256-307x/38/3/037403} {\bibfield  {journal} {\bibinfo  {journal} {Chin. Phys. Lett.}\ }\textbf {\bibinfo {volume} {38}},\ \bibinfo {pages} {037403} (\bibinfo {year} {2021})}\BibitemShut {NoStop}%
\bibitem [{\citenamefont {Yang}\ \emph {et~al.}(2020)\citenamefont {Yang}, \citenamefont {Wang}, \citenamefont {Ortiz}, \citenamefont {Liu}, \citenamefont {Gayles}, \citenamefont {Derunova}, \citenamefont {Gonzalez-Hernandez}, \citenamefont {{\v{s}}mejkal}, \citenamefont {Chen}, \citenamefont {Parkin}, \citenamefont {Wilson}, \citenamefont {Toberer}, \citenamefont {McQueen},\ and\ \citenamefont {Ali}}]{Yang2020}%
  \BibitemOpen
  \bibfield  {author} {\bibinfo {author} {\bibfnamefont {S.-Y.}\ \bibnamefont {Yang}}, \bibinfo {author} {\bibfnamefont {Y.}~\bibnamefont {Wang}}, \bibinfo {author} {\bibfnamefont {B.~R.}\ \bibnamefont {Ortiz}}, \bibinfo {author} {\bibfnamefont {D.}~\bibnamefont {Liu}}, \bibinfo {author} {\bibfnamefont {J.}~\bibnamefont {Gayles}}, \bibinfo {author} {\bibfnamefont {E.}~\bibnamefont {Derunova}}, \bibinfo {author} {\bibfnamefont {R.}~\bibnamefont {Gonzalez-Hernandez}}, \bibinfo {author} {\bibfnamefont {L.}~\bibnamefont {{\v{s}}mejkal}}, \bibinfo {author} {\bibfnamefont {Y.}~\bibnamefont {Chen}}, \bibinfo {author} {\bibfnamefont {S.~S.~P.}\ \bibnamefont {Parkin}}, \bibinfo {author} {\bibfnamefont {S.~D.}\ \bibnamefont {Wilson}}, \bibinfo {author} {\bibfnamefont {E.~S.}\ \bibnamefont {Toberer}}, \bibinfo {author} {\bibfnamefont {T.}~\bibnamefont {McQueen}},\ and\ \bibinfo {author} {\bibfnamefont {M.~N.}\ \bibnamefont {Ali}},\ }\href {https://doi.org/10.1126/sciadv.abb6003} {\bibfield  {journal} {\bibinfo
  {journal} {Sci. Adv.}\ }\textbf {\bibinfo {volume} {6}},\ \bibinfo {pages} {eabb6003} (\bibinfo {year} {2020})}\BibitemShut {NoStop}%
\bibitem [{\citenamefont {Yu}\ \emph {et~al.}(2021{\natexlab{a}})\citenamefont {Yu}, \citenamefont {Wu}, \citenamefont {Wang}, \citenamefont {Lei}, \citenamefont {Zhuo}, \citenamefont {Ying},\ and\ \citenamefont {Chen}}]{Yu2021}%
  \BibitemOpen
  \bibfield  {author} {\bibinfo {author} {\bibfnamefont {F.~H.}\ \bibnamefont {Yu}}, \bibinfo {author} {\bibfnamefont {T.}~\bibnamefont {Wu}}, \bibinfo {author} {\bibfnamefont {Z.~Y.}\ \bibnamefont {Wang}}, \bibinfo {author} {\bibfnamefont {B.}~\bibnamefont {Lei}}, \bibinfo {author} {\bibfnamefont {W.~Z.}\ \bibnamefont {Zhuo}}, \bibinfo {author} {\bibfnamefont {J.~J.}\ \bibnamefont {Ying}},\ and\ \bibinfo {author} {\bibfnamefont {X.~H.}\ \bibnamefont {Chen}},\ }\href {https://doi.org/10.1103/PhysRevB.104.L041103} {\bibfield  {journal} {\bibinfo  {journal} {Phys. Rev. B}\ }\textbf {\bibinfo {volume} {104}},\ \bibinfo {pages} {L041103} (\bibinfo {year} {2021}{\natexlab{a}})}\BibitemShut {NoStop}%
\bibitem [{\citenamefont {Zheng}\ \emph {et~al.}(2023)\citenamefont {Zheng}, \citenamefont {Tan}, \citenamefont {Chen}, \citenamefont {Wang}, \citenamefont {Zhu}, \citenamefont {Albarakati}, \citenamefont {Algarni}, \citenamefont {Partridge}, \citenamefont {Farrar}, \citenamefont {Zhou} \emph {et~al.}}]{Zheng2021_gating}%
  \BibitemOpen
  \bibfield  {author} {\bibinfo {author} {\bibfnamefont {G.}~\bibnamefont {Zheng}}, \bibinfo {author} {\bibfnamefont {C.}~\bibnamefont {Tan}}, \bibinfo {author} {\bibfnamefont {Z.}~\bibnamefont {Chen}}, \bibinfo {author} {\bibfnamefont {M.}~\bibnamefont {Wang}}, \bibinfo {author} {\bibfnamefont {X.}~\bibnamefont {Zhu}}, \bibinfo {author} {\bibfnamefont {S.}~\bibnamefont {Albarakati}}, \bibinfo {author} {\bibfnamefont {M.}~\bibnamefont {Algarni}}, \bibinfo {author} {\bibfnamefont {J.}~\bibnamefont {Partridge}}, \bibinfo {author} {\bibfnamefont {L.}~\bibnamefont {Farrar}}, \bibinfo {author} {\bibfnamefont {J.}~\bibnamefont {Zhou}}, \emph {et~al.},\ }\href {https://doi.org/10.1038/s41467-023-36208-6} {\bibfield  {journal} {\bibinfo  {journal} {Nat. Commun.}\ }\textbf {\bibinfo {volume} {14}},\ \bibinfo {pages} {678} (\bibinfo {year} {2023})}\BibitemShut {NoStop}%
\bibitem [{\citenamefont {Zhou}\ \emph {et~al.}(2022)\citenamefont {Zhou}, \citenamefont {Liu}, \citenamefont {Wu}, \citenamefont {Jiang}, \citenamefont {Shi}, \citenamefont {Li}, \citenamefont {Sui}, \citenamefont {Hu},\ and\ \citenamefont {Luo}}]{Zhou2021}%
  \BibitemOpen
  \bibfield  {author} {\bibinfo {author} {\bibfnamefont {X.}~\bibnamefont {Zhou}}, \bibinfo {author} {\bibfnamefont {H.}~\bibnamefont {Liu}}, \bibinfo {author} {\bibfnamefont {W.}~\bibnamefont {Wu}}, \bibinfo {author} {\bibfnamefont {K.}~\bibnamefont {Jiang}}, \bibinfo {author} {\bibfnamefont {Y.}~\bibnamefont {Shi}}, \bibinfo {author} {\bibfnamefont {Z.}~\bibnamefont {Li}}, \bibinfo {author} {\bibfnamefont {Y.}~\bibnamefont {Sui}}, \bibinfo {author} {\bibfnamefont {J.}~\bibnamefont {Hu}},\ and\ \bibinfo {author} {\bibfnamefont {J.}~\bibnamefont {Luo}},\ }\href {https://doi.org/10.1103/PhysRevB.105.205104} {\bibfield  {journal} {\bibinfo  {journal} {Phys. Rev. B}\ }\textbf {\bibinfo {volume} {105}},\ \bibinfo {pages} {205104} (\bibinfo {year} {2022})}\BibitemShut {NoStop}%
\bibitem [{\citenamefont {Jiang}\ \emph {et~al.}(2021)\citenamefont {Jiang}, \citenamefont {Yin}, \citenamefont {Denner}, \citenamefont {Shumiya}, \citenamefont {Ortiz}, \citenamefont {Xu}, \citenamefont {Guguchia}, \citenamefont {He}, \citenamefont {Hossain}, \citenamefont {Liu}, \citenamefont {Ruff}, \citenamefont {Kautzsch}, \citenamefont {Zhang}, \citenamefont {Chang}, \citenamefont {Belopolski}, \citenamefont {Zhang}, \citenamefont {Cochran}, \citenamefont {Multer}, \citenamefont {Litskevich}, \citenamefont {Cheng}, \citenamefont {Yang}, \citenamefont {Wang}, \citenamefont {Thomale}, \citenamefont {Neupert}, \citenamefont {Wilson},\ and\ \citenamefont {Hasan}}]{Jiang2021a}%
  \BibitemOpen
  \bibfield  {author} {\bibinfo {author} {\bibfnamefont {Y.~X.}\ \bibnamefont {Jiang}}, \bibinfo {author} {\bibfnamefont {J.~X.}\ \bibnamefont {Yin}}, \bibinfo {author} {\bibfnamefont {M.~M.}\ \bibnamefont {Denner}}, \bibinfo {author} {\bibfnamefont {N.}~\bibnamefont {Shumiya}}, \bibinfo {author} {\bibfnamefont {B.~R.}\ \bibnamefont {Ortiz}}, \bibinfo {author} {\bibfnamefont {G.}~\bibnamefont {Xu}}, \bibinfo {author} {\bibfnamefont {Z.}~\bibnamefont {Guguchia}}, \bibinfo {author} {\bibfnamefont {J.}~\bibnamefont {He}}, \bibinfo {author} {\bibfnamefont {M.~S.}\ \bibnamefont {Hossain}}, \bibinfo {author} {\bibfnamefont {X.}~\bibnamefont {Liu}}, \bibinfo {author} {\bibfnamefont {J.}~\bibnamefont {Ruff}}, \bibinfo {author} {\bibfnamefont {L.}~\bibnamefont {Kautzsch}}, \bibinfo {author} {\bibfnamefont {S.~S.}\ \bibnamefont {Zhang}}, \bibinfo {author} {\bibfnamefont {G.}~\bibnamefont {Chang}}, \bibinfo {author} {\bibfnamefont {I.}~\bibnamefont {Belopolski}}, \bibinfo {author} {\bibfnamefont {Q.}~\bibnamefont
  {Zhang}}, \bibinfo {author} {\bibfnamefont {T.~A.}\ \bibnamefont {Cochran}}, \bibinfo {author} {\bibfnamefont {D.}~\bibnamefont {Multer}}, \bibinfo {author} {\bibfnamefont {M.}~\bibnamefont {Litskevich}}, \bibinfo {author} {\bibfnamefont {Z.~J.}\ \bibnamefont {Cheng}}, \bibinfo {author} {\bibfnamefont {X.~P.}\ \bibnamefont {Yang}}, \bibinfo {author} {\bibfnamefont {Z.}~\bibnamefont {Wang}}, \bibinfo {author} {\bibfnamefont {R.}~\bibnamefont {Thomale}}, \bibinfo {author} {\bibfnamefont {T.}~\bibnamefont {Neupert}}, \bibinfo {author} {\bibfnamefont {S.~D.}\ \bibnamefont {Wilson}},\ and\ \bibinfo {author} {\bibfnamefont {M.~Z.}\ \bibnamefont {Hasan}},\ }\href {http://dx.doi.org/10.1038/s41563-021-01034-y} {\bibfield  {journal} {\bibinfo  {journal} {Nat. Mater.}\ }\textbf {\bibinfo {volume} {20}},\ \bibinfo {pages} {1353} (\bibinfo {year} {2021})}\BibitemShut {NoStop}%
\bibitem [{\citenamefont {Wang}\ \emph {et~al.}(2021)\citenamefont {Wang}, \citenamefont {Jiang}, \citenamefont {Yin}, \citenamefont {Li}, \citenamefont {Wang}, \citenamefont {Huang}, \citenamefont {Shao}, \citenamefont {Liu}, \citenamefont {Zhu}, \citenamefont {Shumiya}, \citenamefont {Hossain}, \citenamefont {Liu}, \citenamefont {Shi}, \citenamefont {Duan}, \citenamefont {Li}, \citenamefont {Chang}, \citenamefont {Dai}, \citenamefont {Ye}, \citenamefont {Xu}, \citenamefont {Wang}, \citenamefont {Zheng}, \citenamefont {Jia}, \citenamefont {Hasan},\ and\ \citenamefont {Yao}}]{Wang2021CDW}%
  \BibitemOpen
  \bibfield  {author} {\bibinfo {author} {\bibfnamefont {Z.}~\bibnamefont {Wang}}, \bibinfo {author} {\bibfnamefont {Y.-X.}\ \bibnamefont {Jiang}}, \bibinfo {author} {\bibfnamefont {J.-X.}\ \bibnamefont {Yin}}, \bibinfo {author} {\bibfnamefont {Y.}~\bibnamefont {Li}}, \bibinfo {author} {\bibfnamefont {G.-Y.}\ \bibnamefont {Wang}}, \bibinfo {author} {\bibfnamefont {H.-L.}\ \bibnamefont {Huang}}, \bibinfo {author} {\bibfnamefont {S.}~\bibnamefont {Shao}}, \bibinfo {author} {\bibfnamefont {J.}~\bibnamefont {Liu}}, \bibinfo {author} {\bibfnamefont {P.}~\bibnamefont {Zhu}}, \bibinfo {author} {\bibfnamefont {N.}~\bibnamefont {Shumiya}}, \bibinfo {author} {\bibfnamefont {M.~S.}\ \bibnamefont {Hossain}}, \bibinfo {author} {\bibfnamefont {H.}~\bibnamefont {Liu}}, \bibinfo {author} {\bibfnamefont {Y.}~\bibnamefont {Shi}}, \bibinfo {author} {\bibfnamefont {J.}~\bibnamefont {Duan}}, \bibinfo {author} {\bibfnamefont {X.}~\bibnamefont {Li}}, \bibinfo {author} {\bibfnamefont {G.}~\bibnamefont {Chang}}, \bibinfo {author}
  {\bibfnamefont {P.}~\bibnamefont {Dai}}, \bibinfo {author} {\bibfnamefont {Z.}~\bibnamefont {Ye}}, \bibinfo {author} {\bibfnamefont {G.}~\bibnamefont {Xu}}, \bibinfo {author} {\bibfnamefont {Y.}~\bibnamefont {Wang}}, \bibinfo {author} {\bibfnamefont {H.}~\bibnamefont {Zheng}}, \bibinfo {author} {\bibfnamefont {J.}~\bibnamefont {Jia}}, \bibinfo {author} {\bibfnamefont {M.~Z.}\ \bibnamefont {Hasan}},\ and\ \bibinfo {author} {\bibfnamefont {Y.}~\bibnamefont {Yao}},\ }\href {https://doi.org/10.1103/PhysRevB.104.075148} {\bibfield  {journal} {\bibinfo  {journal} {Phys. Rev. B}\ }\textbf {\bibinfo {volume} {104}},\ \bibinfo {pages} {075148} (\bibinfo {year} {2021})}\BibitemShut {NoStop}%
\bibitem [{\citenamefont {Shumiya}\ \emph {et~al.}(2021)\citenamefont {Shumiya}, \citenamefont {Hossain}, \citenamefont {Yin}, \citenamefont {Jiang}, \citenamefont {Ortiz}, \citenamefont {Liu}, \citenamefont {Shi}, \citenamefont {Yin}, \citenamefont {Lei}, \citenamefont {Zhang}, \citenamefont {Chang}, \citenamefont {Zhang}, \citenamefont {Cochran}, \citenamefont {Multer}, \citenamefont {Litskevich}, \citenamefont {Cheng}, \citenamefont {Yang}, \citenamefont {Guguchia}, \citenamefont {Wilson},\ and\ \citenamefont {Hasan}}]{Shumiya2021}%
  \BibitemOpen
  \bibfield  {author} {\bibinfo {author} {\bibfnamefont {N.}~\bibnamefont {Shumiya}}, \bibinfo {author} {\bibfnamefont {M.~S.}\ \bibnamefont {Hossain}}, \bibinfo {author} {\bibfnamefont {J.-X.}\ \bibnamefont {Yin}}, \bibinfo {author} {\bibfnamefont {Y.-X.}\ \bibnamefont {Jiang}}, \bibinfo {author} {\bibfnamefont {B.~R.}\ \bibnamefont {Ortiz}}, \bibinfo {author} {\bibfnamefont {H.}~\bibnamefont {Liu}}, \bibinfo {author} {\bibfnamefont {Y.}~\bibnamefont {Shi}}, \bibinfo {author} {\bibfnamefont {Q.}~\bibnamefont {Yin}}, \bibinfo {author} {\bibfnamefont {H.}~\bibnamefont {Lei}}, \bibinfo {author} {\bibfnamefont {S.~S.}\ \bibnamefont {Zhang}}, \bibinfo {author} {\bibfnamefont {G.}~\bibnamefont {Chang}}, \bibinfo {author} {\bibfnamefont {Q.}~\bibnamefont {Zhang}}, \bibinfo {author} {\bibfnamefont {T.~A.}\ \bibnamefont {Cochran}}, \bibinfo {author} {\bibfnamefont {D.}~\bibnamefont {Multer}}, \bibinfo {author} {\bibfnamefont {M.}~\bibnamefont {Litskevich}}, \bibinfo {author} {\bibfnamefont {Z.-J.}\ \bibnamefont
  {Cheng}}, \bibinfo {author} {\bibfnamefont {X.~P.}\ \bibnamefont {Yang}}, \bibinfo {author} {\bibfnamefont {Z.}~\bibnamefont {Guguchia}}, \bibinfo {author} {\bibfnamefont {S.~D.}\ \bibnamefont {Wilson}},\ and\ \bibinfo {author} {\bibfnamefont {M.~Z.}\ \bibnamefont {Hasan}},\ }\href {https://doi.org/10.1103/PhysRevB.104.035131} {\bibfield  {journal} {\bibinfo  {journal} {Phys. Rev. B}\ }\textbf {\bibinfo {volume} {104}},\ \bibinfo {pages} {035131} (\bibinfo {year} {2021})}\BibitemShut {NoStop}%
\bibitem [{\citenamefont {Chen}\ \emph {et~al.}(2022)\citenamefont {Chen}, \citenamefont {He}, \citenamefont {Yao}, \citenamefont {Pan}, \citenamefont {Lin}, \citenamefont {Schnelle}, \citenamefont {Sun}, \citenamefont {Gooth}, \citenamefont {Taillefer},\ and\ \citenamefont {Felser}}]{Chendong2021}%
  \BibitemOpen
  \bibfield  {author} {\bibinfo {author} {\bibfnamefont {D.}~\bibnamefont {Chen}}, \bibinfo {author} {\bibfnamefont {B.}~\bibnamefont {He}}, \bibinfo {author} {\bibfnamefont {M.}~\bibnamefont {Yao}}, \bibinfo {author} {\bibfnamefont {Y.}~\bibnamefont {Pan}}, \bibinfo {author} {\bibfnamefont {H.}~\bibnamefont {Lin}}, \bibinfo {author} {\bibfnamefont {W.}~\bibnamefont {Schnelle}}, \bibinfo {author} {\bibfnamefont {Y.}~\bibnamefont {Sun}}, \bibinfo {author} {\bibfnamefont {J.}~\bibnamefont {Gooth}}, \bibinfo {author} {\bibfnamefont {L.}~\bibnamefont {Taillefer}},\ and\ \bibinfo {author} {\bibfnamefont {C.}~\bibnamefont {Felser}},\ }\href {https://doi.org/10.1103/PhysRevB.105.L201109} {\bibfield  {journal} {\bibinfo  {journal} {Phys. Rev. B}\ }\textbf {\bibinfo {volume} {105}},\ \bibinfo {pages} {L201109} (\bibinfo {year} {2022})}\BibitemShut {NoStop}%
\bibitem [{\citenamefont {Mielke}\ \emph {et~al.}(2022)\citenamefont {Mielke}, \citenamefont {Das}, \citenamefont {Yin}, \citenamefont {Liu}, \citenamefont {Gupta}, \citenamefont {Jiang}, \citenamefont {Medarde}, \citenamefont {Wu}, \citenamefont {Lei}, \citenamefont {Chang}, \citenamefont {Dai}, \citenamefont {Si}, \citenamefont {Miao}, \citenamefont {Thomale}, \citenamefont {Neupert}, \citenamefont {Shi}, \citenamefont {Khasanov}, \citenamefont {Hasan}, \citenamefont {Luetkens},\ and\ \citenamefont {Guguchia}}]{Mielke2021}%
  \BibitemOpen
  \bibfield  {author} {\bibinfo {author} {\bibfnamefont {C.}~\bibnamefont {Mielke}}, \bibinfo {author} {\bibfnamefont {D.}~\bibnamefont {Das}}, \bibinfo {author} {\bibfnamefont {J.-X.}\ \bibnamefont {Yin}}, \bibinfo {author} {\bibfnamefont {H.}~\bibnamefont {Liu}}, \bibinfo {author} {\bibfnamefont {R.}~\bibnamefont {Gupta}}, \bibinfo {author} {\bibfnamefont {Y.-X.}\ \bibnamefont {Jiang}}, \bibinfo {author} {\bibfnamefont {M.}~\bibnamefont {Medarde}}, \bibinfo {author} {\bibfnamefont {X.}~\bibnamefont {Wu}}, \bibinfo {author} {\bibfnamefont {H.~C.}\ \bibnamefont {Lei}}, \bibinfo {author} {\bibfnamefont {J.}~\bibnamefont {Chang}}, \bibinfo {author} {\bibfnamefont {P.}~\bibnamefont {Dai}}, \bibinfo {author} {\bibfnamefont {Q.}~\bibnamefont {Si}}, \bibinfo {author} {\bibfnamefont {H.}~\bibnamefont {Miao}}, \bibinfo {author} {\bibfnamefont {R.}~\bibnamefont {Thomale}}, \bibinfo {author} {\bibfnamefont {T.}~\bibnamefont {Neupert}}, \bibinfo {author} {\bibfnamefont {Y.}~\bibnamefont {Shi}}, \bibinfo {author}
  {\bibfnamefont {R.}~\bibnamefont {Khasanov}}, \bibinfo {author} {\bibfnamefont {M.~Z.}\ \bibnamefont {Hasan}}, \bibinfo {author} {\bibfnamefont {H.}~\bibnamefont {Luetkens}},\ and\ \bibinfo {author} {\bibfnamefont {Z.}~\bibnamefont {Guguchia}},\ }\href {https://doi.org/10.1038/s41586-021-04327-z} {\bibfield  {journal} {\bibinfo  {journal} {Nature}\ }\textbf {\bibinfo {volume} {602}},\ \bibinfo {pages} {245} (\bibinfo {year} {2022})}\BibitemShut {NoStop}%
\bibitem [{\citenamefont {Yu}\ \emph {et~al.}(2021{\natexlab{b}})\citenamefont {Yu}, \citenamefont {Wang}, \citenamefont {Zhang}, \citenamefont {Sander}, \citenamefont {Ni}, \citenamefont {Lu}, \citenamefont {Ma}, \citenamefont {Wang}, \citenamefont {Zhao}, \citenamefont {Chen}, \citenamefont {Jiang}, \citenamefont {Zhang}, \citenamefont {Yang}, \citenamefont {Zhou}, \citenamefont {Dong}, \citenamefont {Johnson}, \citenamefont {Graf}, \citenamefont {Hu}, \citenamefont {Gao},\ and\ \citenamefont {Zhao}}]{Yu2021b}%
  \BibitemOpen
  \bibfield  {author} {\bibinfo {author} {\bibfnamefont {L.}~\bibnamefont {Yu}}, \bibinfo {author} {\bibfnamefont {C.}~\bibnamefont {Wang}}, \bibinfo {author} {\bibfnamefont {Y.}~\bibnamefont {Zhang}}, \bibinfo {author} {\bibfnamefont {M.}~\bibnamefont {Sander}}, \bibinfo {author} {\bibfnamefont {S.}~\bibnamefont {Ni}}, \bibinfo {author} {\bibfnamefont {Z.}~\bibnamefont {Lu}}, \bibinfo {author} {\bibfnamefont {S.}~\bibnamefont {Ma}}, \bibinfo {author} {\bibfnamefont {Z.}~\bibnamefont {Wang}}, \bibinfo {author} {\bibfnamefont {Z.}~\bibnamefont {Zhao}}, \bibinfo {author} {\bibfnamefont {H.}~\bibnamefont {Chen}}, \bibinfo {author} {\bibfnamefont {K.}~\bibnamefont {Jiang}}, \bibinfo {author} {\bibfnamefont {Y.}~\bibnamefont {Zhang}}, \bibinfo {author} {\bibfnamefont {H.}~\bibnamefont {Yang}}, \bibinfo {author} {\bibfnamefont {F.}~\bibnamefont {Zhou}}, \bibinfo {author} {\bibfnamefont {X.}~\bibnamefont {Dong}}, \bibinfo {author} {\bibfnamefont {S.~L.}\ \bibnamefont {Johnson}}, \bibinfo {author} {\bibfnamefont
  {M.~J.}\ \bibnamefont {Graf}}, \bibinfo {author} {\bibfnamefont {J.}~\bibnamefont {Hu}}, \bibinfo {author} {\bibfnamefont {H.-J.}\ \bibnamefont {Gao}},\ and\ \bibinfo {author} {\bibfnamefont {Z.}~\bibnamefont {Zhao}},\ }\Eprint {https://arxiv.org/abs/2107.10714} {arXiv:2107.10714}  (\bibinfo {year} {2021}{\natexlab{b}})\BibitemShut {NoStop}%
\bibitem [{\citenamefont {Gan}\ \emph {et~al.}(2021)\citenamefont {Gan}, \citenamefont {Xia}, \citenamefont {Zhang}, \citenamefont {Yang}, \citenamefont {Mi}, \citenamefont {Wang}, \citenamefont {Chai}, \citenamefont {Guo}, \citenamefont {Zhou},\ and\ \citenamefont {He}}]{Gan2021}%
  \BibitemOpen
  \bibfield  {author} {\bibinfo {author} {\bibfnamefont {Y.}~\bibnamefont {Gan}}, \bibinfo {author} {\bibfnamefont {W.}~\bibnamefont {Xia}}, \bibinfo {author} {\bibfnamefont {L.}~\bibnamefont {Zhang}}, \bibinfo {author} {\bibfnamefont {K.}~\bibnamefont {Yang}}, \bibinfo {author} {\bibfnamefont {X.}~\bibnamefont {Mi}}, \bibinfo {author} {\bibfnamefont {A.}~\bibnamefont {Wang}}, \bibinfo {author} {\bibfnamefont {Y.}~\bibnamefont {Chai}}, \bibinfo {author} {\bibfnamefont {Y.}~\bibnamefont {Guo}}, \bibinfo {author} {\bibfnamefont {X.}~\bibnamefont {Zhou}},\ and\ \bibinfo {author} {\bibfnamefont {M.}~\bibnamefont {He}},\ }\href {https://doi.org/10.1103/PhysRevB.104.L180508} {\bibfield  {journal} {\bibinfo  {journal} {Phys. Rev. B}\ }\textbf {\bibinfo {volume} {104}},\ \bibinfo {pages} {L180508} (\bibinfo {year} {2021})}\BibitemShut {NoStop}%
\bibitem [{\citenamefont {Mi}\ \emph {et~al.}(2022)\citenamefont {Mi}, \citenamefont {Xia}, \citenamefont {Zhang}, \citenamefont {Gan}, \citenamefont {Yang}, \citenamefont {Wang}, \citenamefont {Chai}, \citenamefont {Guo}, \citenamefont {Zhou},\ and\ \citenamefont {He}}]{Mi_2022}%
  \BibitemOpen
  \bibfield  {author} {\bibinfo {author} {\bibfnamefont {X.}~\bibnamefont {Mi}}, \bibinfo {author} {\bibfnamefont {W.}~\bibnamefont {Xia}}, \bibinfo {author} {\bibfnamefont {L.}~\bibnamefont {Zhang}}, \bibinfo {author} {\bibfnamefont {Y.}~\bibnamefont {Gan}}, \bibinfo {author} {\bibfnamefont {K.}~\bibnamefont {Yang}}, \bibinfo {author} {\bibfnamefont {A.}~\bibnamefont {Wang}}, \bibinfo {author} {\bibfnamefont {Y.}~\bibnamefont {Chai}}, \bibinfo {author} {\bibfnamefont {Y.}~\bibnamefont {Guo}}, \bibinfo {author} {\bibfnamefont {X.}~\bibnamefont {Zhou}},\ and\ \bibinfo {author} {\bibfnamefont {M.}~\bibnamefont {He}},\ }\href {https://doi.org/10.1088/1367-2630/ac8e24} {\bibfield  {journal} {\bibinfo  {journal} {New J. Phys.}\ }\textbf {\bibinfo {volume} {24}},\ \bibinfo {pages} {093021} (\bibinfo {year} {2022})}\BibitemShut {NoStop}%
\bibitem [{\citenamefont {Liang}\ \emph {et~al.}(2021)\citenamefont {Liang}, \citenamefont {Hou}, \citenamefont {Zhang}, \citenamefont {Ma}, \citenamefont {Wu}, \citenamefont {Zhang}, \citenamefont {Yu}, \citenamefont {Ying}, \citenamefont {Jiang}, \citenamefont {Shan}, \citenamefont {Wang},\ and\ \citenamefont {Chen}}]{Liang2021CDW}%
  \BibitemOpen
  \bibfield  {author} {\bibinfo {author} {\bibfnamefont {Z.}~\bibnamefont {Liang}}, \bibinfo {author} {\bibfnamefont {X.}~\bibnamefont {Hou}}, \bibinfo {author} {\bibfnamefont {F.}~\bibnamefont {Zhang}}, \bibinfo {author} {\bibfnamefont {W.}~\bibnamefont {Ma}}, \bibinfo {author} {\bibfnamefont {P.}~\bibnamefont {Wu}}, \bibinfo {author} {\bibfnamefont {Z.}~\bibnamefont {Zhang}}, \bibinfo {author} {\bibfnamefont {F.}~\bibnamefont {Yu}}, \bibinfo {author} {\bibfnamefont {J.-J.}\ \bibnamefont {Ying}}, \bibinfo {author} {\bibfnamefont {K.}~\bibnamefont {Jiang}}, \bibinfo {author} {\bibfnamefont {L.}~\bibnamefont {Shan}}, \bibinfo {author} {\bibfnamefont {Z.}~\bibnamefont {Wang}},\ and\ \bibinfo {author} {\bibfnamefont {X.-H.}\ \bibnamefont {Chen}},\ }\href {https://doi.org/10.1103/PhysRevX.11.031026} {\bibfield  {journal} {\bibinfo  {journal} {Phys. Rev. X}\ }\textbf {\bibinfo {volume} {11}},\ \bibinfo {pages} {031026} (\bibinfo {year} {2021})}\BibitemShut {NoStop}%
\bibitem [{\citenamefont {Zhao}\ \emph {et~al.}(2021)\citenamefont {Zhao}, \citenamefont {Li}, \citenamefont {Ortiz}, \citenamefont {Teicher}, \citenamefont {Park}, \citenamefont {Ye}, \citenamefont {Wang}, \citenamefont {Balents}, \citenamefont {Wilson},\ and\ \citenamefont {Zeljkovic}}]{Zhao2021a}%
  \BibitemOpen
  \bibfield  {author} {\bibinfo {author} {\bibfnamefont {H.}~\bibnamefont {Zhao}}, \bibinfo {author} {\bibfnamefont {H.}~\bibnamefont {Li}}, \bibinfo {author} {\bibfnamefont {B.~R.}\ \bibnamefont {Ortiz}}, \bibinfo {author} {\bibfnamefont {S.~M.~L.}\ \bibnamefont {Teicher}}, \bibinfo {author} {\bibfnamefont {T.}~\bibnamefont {Park}}, \bibinfo {author} {\bibfnamefont {M.}~\bibnamefont {Ye}}, \bibinfo {author} {\bibfnamefont {Z.}~\bibnamefont {Wang}}, \bibinfo {author} {\bibfnamefont {L.}~\bibnamefont {Balents}}, \bibinfo {author} {\bibfnamefont {S.~D.}\ \bibnamefont {Wilson}},\ and\ \bibinfo {author} {\bibfnamefont {I.}~\bibnamefont {Zeljkovic}},\ }\href {https://doi.org/10.1038/s41586-021-03946-w} {\bibfield  {journal} {\bibinfo  {journal} {Nature}\ }\textbf {\bibinfo {volume} {599}},\ \bibinfo {pages} {216} (\bibinfo {year} {2021})}\BibitemShut {NoStop}%
\bibitem [{\citenamefont {Chen}\ \emph {et~al.}(2021)\citenamefont {Chen}, \citenamefont {Yang}, \citenamefont {Hu}, \citenamefont {Zhao}, \citenamefont {Yuan}, \citenamefont {Xing}, \citenamefont {Qian}, \citenamefont {Huang}, \citenamefont {Li}, \citenamefont {Ye}, \citenamefont {Ma}, \citenamefont {Ni}, \citenamefont {Zhang}, \citenamefont {Yin}, \citenamefont {Gong}, \citenamefont {Tu}, \citenamefont {Lei}, \citenamefont {Tan}, \citenamefont {Zhou}, \citenamefont {Shen}, \citenamefont {Dong}, \citenamefont {Yan}, \citenamefont {Wang},\ and\ \citenamefont {Gao}}]{Chen2021rotonpair}%
  \BibitemOpen
  \bibfield  {author} {\bibinfo {author} {\bibfnamefont {H.}~\bibnamefont {Chen}}, \bibinfo {author} {\bibfnamefont {H.}~\bibnamefont {Yang}}, \bibinfo {author} {\bibfnamefont {B.}~\bibnamefont {Hu}}, \bibinfo {author} {\bibfnamefont {Z.}~\bibnamefont {Zhao}}, \bibinfo {author} {\bibfnamefont {J.}~\bibnamefont {Yuan}}, \bibinfo {author} {\bibfnamefont {Y.}~\bibnamefont {Xing}}, \bibinfo {author} {\bibfnamefont {G.}~\bibnamefont {Qian}}, \bibinfo {author} {\bibfnamefont {Z.}~\bibnamefont {Huang}}, \bibinfo {author} {\bibfnamefont {G.}~\bibnamefont {Li}}, \bibinfo {author} {\bibfnamefont {Y.}~\bibnamefont {Ye}}, \bibinfo {author} {\bibfnamefont {S.}~\bibnamefont {Ma}}, \bibinfo {author} {\bibfnamefont {S.}~\bibnamefont {Ni}}, \bibinfo {author} {\bibfnamefont {H.}~\bibnamefont {Zhang}}, \bibinfo {author} {\bibfnamefont {Q.}~\bibnamefont {Yin}}, \bibinfo {author} {\bibfnamefont {C.}~\bibnamefont {Gong}}, \bibinfo {author} {\bibfnamefont {Z.}~\bibnamefont {Tu}}, \bibinfo {author} {\bibfnamefont {H.}~\bibnamefont
  {Lei}}, \bibinfo {author} {\bibfnamefont {H.}~\bibnamefont {Tan}}, \bibinfo {author} {\bibfnamefont {S.}~\bibnamefont {Zhou}}, \bibinfo {author} {\bibfnamefont {C.}~\bibnamefont {Shen}}, \bibinfo {author} {\bibfnamefont {X.}~\bibnamefont {Dong}}, \bibinfo {author} {\bibfnamefont {B.}~\bibnamefont {Yan}}, \bibinfo {author} {\bibfnamefont {Z.}~\bibnamefont {Wang}},\ and\ \bibinfo {author} {\bibfnamefont {H.-J.}\ \bibnamefont {Gao}},\ }\href {https://doi.org/10.1038/s41586-021-03983-5} {\bibfield  {journal} {\bibinfo  {journal} {Nature}\ }\textbf {\bibinfo {volume} {599}},\ \bibinfo {pages} {222} (\bibinfo {year} {2021})}\BibitemShut {NoStop}%
\bibitem [{\citenamefont {Xu}\ \emph {et~al.}(2021{\natexlab{a}})\citenamefont {Xu}, \citenamefont {Yan}, \citenamefont {Yin}, \citenamefont {Xia}, \citenamefont {Fang}, \citenamefont {Chen}, \citenamefont {Li}, \citenamefont {Yang}, \citenamefont {Guo},\ and\ \citenamefont {Feng}}]{Xu2021}%
  \BibitemOpen
  \bibfield  {author} {\bibinfo {author} {\bibfnamefont {H.-S.}\ \bibnamefont {Xu}}, \bibinfo {author} {\bibfnamefont {Y.-J.}\ \bibnamefont {Yan}}, \bibinfo {author} {\bibfnamefont {R.}~\bibnamefont {Yin}}, \bibinfo {author} {\bibfnamefont {W.}~\bibnamefont {Xia}}, \bibinfo {author} {\bibfnamefont {S.}~\bibnamefont {Fang}}, \bibinfo {author} {\bibfnamefont {Z.}~\bibnamefont {Chen}}, \bibinfo {author} {\bibfnamefont {Y.}~\bibnamefont {Li}}, \bibinfo {author} {\bibfnamefont {W.}~\bibnamefont {Yang}}, \bibinfo {author} {\bibfnamefont {Y.}~\bibnamefont {Guo}},\ and\ \bibinfo {author} {\bibfnamefont {D.-L.}\ \bibnamefont {Feng}},\ }\href {https://doi.org/10.1103/PhysRevLett.127.187004} {\bibfield  {journal} {\bibinfo  {journal} {Phys. Rev. Lett.}\ }\textbf {\bibinfo {volume} {127}},\ \bibinfo {pages} {187004} (\bibinfo {year} {2021}{\natexlab{a}})}\BibitemShut {NoStop}%
\bibitem [{\citenamefont {Li}\ \emph {et~al.}(2022)\citenamefont {Li}, \citenamefont {Zhao}, \citenamefont {Ortiz}, \citenamefont {Park}, \citenamefont {Ye}, \citenamefont {Balents}, \citenamefont {Wang}, \citenamefont {Wilson},\ and\ \citenamefont {Zeljkovic}}]{Li2021a}%
  \BibitemOpen
  \bibfield  {author} {\bibinfo {author} {\bibfnamefont {H.}~\bibnamefont {Li}}, \bibinfo {author} {\bibfnamefont {H.}~\bibnamefont {Zhao}}, \bibinfo {author} {\bibfnamefont {B.~R.}\ \bibnamefont {Ortiz}}, \bibinfo {author} {\bibfnamefont {T.}~\bibnamefont {Park}}, \bibinfo {author} {\bibfnamefont {M.}~\bibnamefont {Ye}}, \bibinfo {author} {\bibfnamefont {L.}~\bibnamefont {Balents}}, \bibinfo {author} {\bibfnamefont {Z.}~\bibnamefont {Wang}}, \bibinfo {author} {\bibfnamefont {S.~D.}\ \bibnamefont {Wilson}},\ and\ \bibinfo {author} {\bibfnamefont {I.}~\bibnamefont {Zeljkovic}},\ }\href {https://doi.org/10.1038/s41567-021-01479-7} {\bibfield  {journal} {\bibinfo  {journal} {Nat. Phys.}\ }\textbf {\bibinfo {volume} {18}},\ \bibinfo {pages} {265} (\bibinfo {year} {2022})}\BibitemShut {NoStop}%
\bibitem [{\citenamefont {Hu}\ \emph {et~al.}(2022)\citenamefont {Hu}, \citenamefont {Wu}, \citenamefont {Ortiz}, \citenamefont {Han}, \citenamefont {Plumb}, \citenamefont {Wilson}, \citenamefont {Schnyder},\ and\ \citenamefont {Shi}}]{Hu2022co}%
  \BibitemOpen
  \bibfield  {author} {\bibinfo {author} {\bibfnamefont {Y.}~\bibnamefont {Hu}}, \bibinfo {author} {\bibfnamefont {X.}~\bibnamefont {Wu}}, \bibinfo {author} {\bibfnamefont {B.~R.}\ \bibnamefont {Ortiz}}, \bibinfo {author} {\bibfnamefont {X.}~\bibnamefont {Han}}, \bibinfo {author} {\bibfnamefont {N.~C.}\ \bibnamefont {Plumb}}, \bibinfo {author} {\bibfnamefont {S.~D.}\ \bibnamefont {Wilson}}, \bibinfo {author} {\bibfnamefont {A.~P.}\ \bibnamefont {Schnyder}},\ and\ \bibinfo {author} {\bibfnamefont {M.}~\bibnamefont {Shi}},\ }\href {https://doi.org/10.1103/PhysRevB.106.L241106} {\bibfield  {journal} {\bibinfo  {journal} {Phys. Rev. B}\ }\textbf {\bibinfo {volume} {106}},\ \bibinfo {pages} {L241106} (\bibinfo {year} {2022})}\BibitemShut {NoStop}%
\bibitem [{\citenamefont {Mi}\ \emph {et~al.}(2023)\citenamefont {Mi}, \citenamefont {Yang}, \citenamefont {Gan}, \citenamefont {Zhang}, \citenamefont {Wang}, \citenamefont {Chai}, \citenamefont {Zhou},\ and\ \citenamefont {He}}]{Mi2023}%
  \BibitemOpen
  \bibfield  {author} {\bibinfo {author} {\bibfnamefont {X.}~\bibnamefont {Mi}}, \bibinfo {author} {\bibfnamefont {K.}~\bibnamefont {Yang}}, \bibinfo {author} {\bibfnamefont {Y.}~\bibnamefont {Gan}}, \bibinfo {author} {\bibfnamefont {L.}~\bibnamefont {Zhang}}, \bibinfo {author} {\bibfnamefont {A.}~\bibnamefont {Wang}}, \bibinfo {author} {\bibfnamefont {Y.}~\bibnamefont {Chai}}, \bibinfo {author} {\bibfnamefont {X.}~\bibnamefont {Zhou}},\ and\ \bibinfo {author} {\bibfnamefont {M.}~\bibnamefont {He}},\ }\href {https://doi.org/10.1007/s42864-022-00192-z} {\bibfield  {journal} {\bibinfo  {journal} {Tungsten}\ }\textbf {\bibinfo {volume} {5}},\ \bibinfo {pages} {300} (\bibinfo {year} {2023})}\BibitemShut {NoStop}%
\bibitem [{\citenamefont {Koshelev}\ \emph {et~al.}(2024)\citenamefont {Koshelev}, \citenamefont {Chapai}, \citenamefont {Chung}, \citenamefont {Mitchell},\ and\ \citenamefont {Welp}}]{Koshelev}%
  \BibitemOpen
  \bibfield  {author} {\bibinfo {author} {\bibfnamefont {A.~E.}\ \bibnamefont {Koshelev}}, \bibinfo {author} {\bibfnamefont {R.}~\bibnamefont {Chapai}}, \bibinfo {author} {\bibfnamefont {D.~Y.}\ \bibnamefont {Chung}}, \bibinfo {author} {\bibfnamefont {J.~F.}\ \bibnamefont {Mitchell}},\ and\ \bibinfo {author} {\bibfnamefont {U.}~\bibnamefont {Welp}},\ }\href {https://doi.org/10.1103/PhysRevB.110.024512} {\bibfield  {journal} {\bibinfo  {journal} {Phys. Rev. B}\ }\textbf {\bibinfo {volume} {110}},\ \bibinfo {pages} {024512} (\bibinfo {year} {2024})}\BibitemShut {NoStop}%
\bibitem [{\citenamefont {Kohn}\ and\ \citenamefont {Sham}(1965)}]{Kohn1965}%
  \BibitemOpen
  \bibfield  {author} {\bibinfo {author} {\bibfnamefont {W.}~\bibnamefont {Kohn}}\ and\ \bibinfo {author} {\bibfnamefont {L.~J.}\ \bibnamefont {Sham}},\ }\href {https://doi.org/10.1103/PhysRev.140.A1133} {\bibfield  {journal} {\bibinfo  {journal} {Phys. Rev.}\ }\textbf {\bibinfo {volume} {140}},\ \bibinfo {pages} {A1133} (\bibinfo {year} {1965})}\BibitemShut {NoStop}%
\bibitem [{\citenamefont {Bl\"ochl}(1994)}]{PhysRevB.50.17953}%
  \BibitemOpen
  \bibfield  {author} {\bibinfo {author} {\bibfnamefont {P.~E.}\ \bibnamefont {Bl\"ochl}},\ }\href {https://doi.org/10.1103/PhysRevB.50.17953} {\bibfield  {journal} {\bibinfo  {journal} {Phys. Rev. B}\ }\textbf {\bibinfo {volume} {50}},\ \bibinfo {pages} {17953} (\bibinfo {year} {1994})}\BibitemShut {NoStop}%
\bibitem [{\citenamefont {Kresse}\ and\ \citenamefont {Furthm\"uller}(1996)}]{PhysRevB.54.11169}%
  \BibitemOpen
  \bibfield  {author} {\bibinfo {author} {\bibfnamefont {G.}~\bibnamefont {Kresse}}\ and\ \bibinfo {author} {\bibfnamefont {J.}~\bibnamefont {Furthm\"uller}},\ }\href {https://doi.org/10.1103/PhysRevB.54.11169} {\bibfield  {journal} {\bibinfo  {journal} {Phys. Rev. B}\ }\textbf {\bibinfo {volume} {54}},\ \bibinfo {pages} {11169} (\bibinfo {year} {1996})}\BibitemShut {NoStop}%
\bibitem [{\citenamefont {Perdew}\ \emph {et~al.}(1996)\citenamefont {Perdew}, \citenamefont {Burke},\ and\ \citenamefont {Ernzerhof}}]{PhysRevLett.77.3865}%
  \BibitemOpen
  \bibfield  {author} {\bibinfo {author} {\bibfnamefont {J.~P.}\ \bibnamefont {Perdew}}, \bibinfo {author} {\bibfnamefont {K.}~\bibnamefont {Burke}},\ and\ \bibinfo {author} {\bibfnamefont {M.}~\bibnamefont {Ernzerhof}},\ }\href {https://doi.org/10.1103/PhysRevLett.77.3865} {\bibfield  {journal} {\bibinfo  {journal} {Phys. Rev. Lett.}\ }\textbf {\bibinfo {volume} {77}},\ \bibinfo {pages} {3865} (\bibinfo {year} {1996})}\BibitemShut {NoStop}%
\bibitem [{\citenamefont {Kulik}\ \emph {et~al.}(2006)\citenamefont {Kulik}, \citenamefont {Cococcioni}, \citenamefont {Scherlis},\ and\ \citenamefont {Marzari}}]{PhysRevLett.97.103001}%
  \BibitemOpen
  \bibfield  {author} {\bibinfo {author} {\bibfnamefont {H.~J.}\ \bibnamefont {Kulik}}, \bibinfo {author} {\bibfnamefont {M.}~\bibnamefont {Cococcioni}}, \bibinfo {author} {\bibfnamefont {D.~A.}\ \bibnamefont {Scherlis}},\ and\ \bibinfo {author} {\bibfnamefont {N.}~\bibnamefont {Marzari}},\ }\href {https://doi.org/10.1103/PhysRevLett.97.103001} {\bibfield  {journal} {\bibinfo  {journal} {Phys. Rev. Lett.}\ }\textbf {\bibinfo {volume} {97}},\ \bibinfo {pages} {103001} (\bibinfo {year} {2006})}\BibitemShut {NoStop}%
\bibitem [{\citenamefont {Guo}\ \emph {et~al.}(2024)\citenamefont {Guo}, \citenamefont {Zhou}, \citenamefont {Ding}, \citenamefont {Qu}, \citenamefont {Liu}, \citenamefont {Du}, \citenamefont {Zhang}, \citenamefont {Li}, \citenamefont {Zhang}, \citenamefont {Zhou}, \citenamefont {Qi}, \citenamefont {Cui}, \citenamefont {Zhang}, \citenamefont {Guo}, \citenamefont {Wang}, \citenamefont {Fei}, \citenamefont {Huang}, \citenamefont {Qian}, \citenamefont {Shen}, \citenamefont {Song}, \citenamefont {Weng},\ and\ \citenamefont {Song}}]{Guo_EuTiBi}%
  \BibitemOpen
  \bibfield  {author} {\bibinfo {author} {\bibfnamefont {J.}~\bibnamefont {Guo}}, \bibinfo {author} {\bibfnamefont {L.}~\bibnamefont {Zhou}}, \bibinfo {author} {\bibfnamefont {J.}~\bibnamefont {Ding}}, \bibinfo {author} {\bibfnamefont {G.}~\bibnamefont {Qu}}, \bibinfo {author} {\bibfnamefont {Z.}~\bibnamefont {Liu}}, \bibinfo {author} {\bibfnamefont {Y.}~\bibnamefont {Du}}, \bibinfo {author} {\bibfnamefont {H.}~\bibnamefont {Zhang}}, \bibinfo {author} {\bibfnamefont {J.}~\bibnamefont {Li}}, \bibinfo {author} {\bibfnamefont {Y.}~\bibnamefont {Zhang}}, \bibinfo {author} {\bibfnamefont {F.}~\bibnamefont {Zhou}}, \bibinfo {author} {\bibfnamefont {W.}~\bibnamefont {Qi}}, \bibinfo {author} {\bibfnamefont {M.}~\bibnamefont {Cui}}, \bibinfo {author} {\bibfnamefont {Y.}~\bibnamefont {Zhang}}, \bibinfo {author} {\bibfnamefont {F.}~\bibnamefont {Guo}}, \bibinfo {author} {\bibfnamefont {T.}~\bibnamefont {Wang}}, \bibinfo {author} {\bibfnamefont {F.}~\bibnamefont {Fei}}, \bibinfo {author} {\bibfnamefont {Y.}~\bibnamefont
  {Huang}}, \bibinfo {author} {\bibfnamefont {T.}~\bibnamefont {Qian}}, \bibinfo {author} {\bibfnamefont {D.}~\bibnamefont {Shen}}, \bibinfo {author} {\bibfnamefont {Y.}~\bibnamefont {Song}}, \bibinfo {author} {\bibfnamefont {H.}~\bibnamefont {Weng}},\ and\ \bibinfo {author} {\bibfnamefont {F.}~\bibnamefont {Song}},\ }\href {https://doi.org/https://doi.org/10.1016/j.scib.2024.06.036} {\bibfield  {journal} {\bibinfo  {journal} {Sci. Bull.}\ }\textbf {\bibinfo {volume} {69}},\ \bibinfo {pages} {2660} (\bibinfo {year} {2024})}\BibitemShut {NoStop}%
\bibitem [{\citenamefont {Mostofi}\ \emph {et~al.}(2014)\citenamefont {Mostofi}, \citenamefont {Yates}, \citenamefont {Pizzi}, \citenamefont {Lee}, \citenamefont {Souza}, \citenamefont {Vanderbilt},\ and\ \citenamefont {Marzari}}]{mostofi2014updated}%
  \BibitemOpen
  \bibfield  {author} {\bibinfo {author} {\bibfnamefont {A.~A.}\ \bibnamefont {Mostofi}}, \bibinfo {author} {\bibfnamefont {J.~R.}\ \bibnamefont {Yates}}, \bibinfo {author} {\bibfnamefont {G.}~\bibnamefont {Pizzi}}, \bibinfo {author} {\bibfnamefont {Y.-S.}\ \bibnamefont {Lee}}, \bibinfo {author} {\bibfnamefont {I.}~\bibnamefont {Souza}}, \bibinfo {author} {\bibfnamefont {D.}~\bibnamefont {Vanderbilt}},\ and\ \bibinfo {author} {\bibfnamefont {N.}~\bibnamefont {Marzari}},\ }\href {https://doi.org/10.1016/j.cpc.2014.05.003} {\bibfield  {journal} {\bibinfo  {journal} {Comput. Phys. Commun.}\ }\textbf {\bibinfo {volume} {185}},\ \bibinfo {pages} {2309} (\bibinfo {year} {2014})}\BibitemShut {NoStop}%
\bibitem [{\citenamefont {Wu}\ \emph {et~al.}(2018)\citenamefont {Wu}, \citenamefont {Zhang}, \citenamefont {Song}, \citenamefont {Troyer},\ and\ \citenamefont {Soluyanov}}]{wu2018wanniertools}%
  \BibitemOpen
  \bibfield  {author} {\bibinfo {author} {\bibfnamefont {Q.}~\bibnamefont {Wu}}, \bibinfo {author} {\bibfnamefont {S.}~\bibnamefont {Zhang}}, \bibinfo {author} {\bibfnamefont {H.-F.}\ \bibnamefont {Song}}, \bibinfo {author} {\bibfnamefont {M.}~\bibnamefont {Troyer}},\ and\ \bibinfo {author} {\bibfnamefont {A.~A.}\ \bibnamefont {Soluyanov}},\ }\href {https://doi.org/10.1016/j.cpc.2017.09.033} {\bibfield  {journal} {\bibinfo  {journal} {Comput. Phys. Commun.}\ }\textbf {\bibinfo {volume} {224}},\ \bibinfo {pages} {405} (\bibinfo {year} {2018})}\BibitemShut {NoStop}%
\bibitem [{\citenamefont {Ortiz}\ \emph {et~al.}(2023)\citenamefont {Ortiz}, \citenamefont {Miao}, \citenamefont {Parker}, \citenamefont {Yang}, \citenamefont {Samolyuk}, \citenamefont {Clements}, \citenamefont {Rajapitamahuni}, \citenamefont {Yilmaz}, \citenamefont {Vescovo}, \citenamefont {Yan}, \citenamefont {May},\ and\ \citenamefont {McGuire}}]{Ortiz_EuTiBi}%
  \BibitemOpen
  \bibfield  {author} {\bibinfo {author} {\bibfnamefont {B.~R.}\ \bibnamefont {Ortiz}}, \bibinfo {author} {\bibfnamefont {H.}~\bibnamefont {Miao}}, \bibinfo {author} {\bibfnamefont {D.~S.}\ \bibnamefont {Parker}}, \bibinfo {author} {\bibfnamefont {F.}~\bibnamefont {Yang}}, \bibinfo {author} {\bibfnamefont {G.~D.}\ \bibnamefont {Samolyuk}}, \bibinfo {author} {\bibfnamefont {E.~M.}\ \bibnamefont {Clements}}, \bibinfo {author} {\bibfnamefont {A.}~\bibnamefont {Rajapitamahuni}}, \bibinfo {author} {\bibfnamefont {T.}~\bibnamefont {Yilmaz}}, \bibinfo {author} {\bibfnamefont {E.}~\bibnamefont {Vescovo}}, \bibinfo {author} {\bibfnamefont {J.}~\bibnamefont {Yan}}, \bibinfo {author} {\bibfnamefont {A.~F.}\ \bibnamefont {May}},\ and\ \bibinfo {author} {\bibfnamefont {M.~A.}\ \bibnamefont {McGuire}},\ }\href {https://doi.org/10.1021/acs.chemmater.3c02289} {\bibfield  {journal} {\bibinfo  {journal} {Chem. Mater.}\ }\textbf {\bibinfo {volume} {35}},\ \bibinfo {pages} {9756} (\bibinfo {year} {2023})}\BibitemShut {NoStop}%
\bibitem [{\citenamefont {Jiang}\ \emph {et~al.}(2024)\citenamefont {Jiang}, \citenamefont {Li}, \citenamefont {Yuan}, \citenamefont {Liu}, \citenamefont {Cao}, \citenamefont {Cho}, \citenamefont {Shu}, \citenamefont {Yang}, \citenamefont {Li}, \citenamefont {Liu}, \citenamefont {Ding}, \citenamefont {Liu}, \citenamefont {Liu}, \citenamefont {Ma}, \citenamefont {Sun}, \citenamefont {Wan}, \citenamefont {Guo}, \citenamefont {Shen},\ and\ \citenamefont {Feng}}]{Jiang_EuTiBI}%
  \BibitemOpen
  \bibfield  {author} {\bibinfo {author} {\bibfnamefont {Z.}~\bibnamefont {Jiang}}, \bibinfo {author} {\bibfnamefont {T.}~\bibnamefont {Li}}, \bibinfo {author} {\bibfnamefont {J.}~\bibnamefont {Yuan}}, \bibinfo {author} {\bibfnamefont {Z.}~\bibnamefont {Liu}}, \bibinfo {author} {\bibfnamefont {Z.}~\bibnamefont {Cao}}, \bibinfo {author} {\bibfnamefont {S.}~\bibnamefont {Cho}}, \bibinfo {author} {\bibfnamefont {M.}~\bibnamefont {Shu}}, \bibinfo {author} {\bibfnamefont {Y.}~\bibnamefont {Yang}}, \bibinfo {author} {\bibfnamefont {Z.}~\bibnamefont {Li}}, \bibinfo {author} {\bibfnamefont {J.}~\bibnamefont {Liu}}, \bibinfo {author} {\bibfnamefont {J.}~\bibnamefont {Ding}}, \bibinfo {author} {\bibfnamefont {Z.}~\bibnamefont {Liu}}, \bibinfo {author} {\bibfnamefont {J.}~\bibnamefont {Liu}}, \bibinfo {author} {\bibfnamefont {J.}~\bibnamefont {Ma}}, \bibinfo {author} {\bibfnamefont {Z.}~\bibnamefont {Sun}}, \bibinfo {author} {\bibfnamefont {X.}~\bibnamefont {Wan}}, \bibinfo {author} {\bibfnamefont {Y.}~\bibnamefont
  {Guo}}, \bibinfo {author} {\bibfnamefont {D.}~\bibnamefont {Shen}},\ and\ \bibinfo {author} {\bibfnamefont {D.}~\bibnamefont {Feng}},\ }\href {https://doi.org/https://doi.org/10.1016/j.scib.2024.08.019} {\bibfield  {journal} {\bibinfo  {journal} {Sci. Bull.}\ }\textbf {\bibinfo {volume} {69}},\ \bibinfo {pages} {3192} (\bibinfo {year} {2024})}\BibitemShut {NoStop}%
\bibitem [{\citenamefont {Chen}\ \emph {et~al.}(2020)\citenamefont {Chen}, \citenamefont {Lou}, \citenamefont {Zhou}, \citenamefont {Chen}, \citenamefont {Xu}, \citenamefont {Chen}, \citenamefont {Du}, \citenamefont {Yang}, \citenamefont {Wang},\ and\ \citenamefont {Fang}}]{Chen_2020}%
  \BibitemOpen
  \bibfield  {author} {\bibinfo {author} {\bibfnamefont {H.-C.}\ \bibnamefont {Chen}}, \bibinfo {author} {\bibfnamefont {Z.-F.}\ \bibnamefont {Lou}}, \bibinfo {author} {\bibfnamefont {Y.-X.}\ \bibnamefont {Zhou}}, \bibinfo {author} {\bibfnamefont {Q.}~\bibnamefont {Chen}}, \bibinfo {author} {\bibfnamefont {B.-J.}\ \bibnamefont {Xu}}, \bibinfo {author} {\bibfnamefont {S.-J.}\ \bibnamefont {Chen}}, \bibinfo {author} {\bibfnamefont {J.-H.}\ \bibnamefont {Du}}, \bibinfo {author} {\bibfnamefont {J.-H.}\ \bibnamefont {Yang}}, \bibinfo {author} {\bibfnamefont {H.-D.}\ \bibnamefont {Wang}},\ and\ \bibinfo {author} {\bibfnamefont {M.-H.}\ \bibnamefont {Fang}},\ }\href {https://doi.org/10.1088/0256-307X/37/4/047201} {\bibfield  {journal} {\bibinfo  {journal} {Chin. Phys. Lett.}\ }\textbf {\bibinfo {volume} {37}},\ \bibinfo {pages} {047201} (\bibinfo {year} {2020})}\BibitemShut {NoStop}%
\bibitem [{\citenamefont {Gui}\ \emph {et~al.}(2019)\citenamefont {Gui}, \citenamefont {Pletikosic}, \citenamefont {Cao}, \citenamefont {Tien}, \citenamefont {Xu}, \citenamefont {Zhong}, \citenamefont {Wang}, \citenamefont {Chang}, \citenamefont {Jia}, \citenamefont {Valla}, \citenamefont {Xie},\ and\ \citenamefont {Cava}}]{xinEuSn2P2}%
  \BibitemOpen
  \bibfield  {author} {\bibinfo {author} {\bibfnamefont {X.}~\bibnamefont {Gui}}, \bibinfo {author} {\bibfnamefont {I.}~\bibnamefont {Pletikosic}}, \bibinfo {author} {\bibfnamefont {H.}~\bibnamefont {Cao}}, \bibinfo {author} {\bibfnamefont {H.-J.}\ \bibnamefont {Tien}}, \bibinfo {author} {\bibfnamefont {X.}~\bibnamefont {Xu}}, \bibinfo {author} {\bibfnamefont {R.}~\bibnamefont {Zhong}}, \bibinfo {author} {\bibfnamefont {G.}~\bibnamefont {Wang}}, \bibinfo {author} {\bibfnamefont {T.-R.}\ \bibnamefont {Chang}}, \bibinfo {author} {\bibfnamefont {S.}~\bibnamefont {Jia}}, \bibinfo {author} {\bibfnamefont {T.}~\bibnamefont {Valla}}, \bibinfo {author} {\bibfnamefont {W.}~\bibnamefont {Xie}},\ and\ \bibinfo {author} {\bibfnamefont {R.~J.}\ \bibnamefont {Cava}},\ }\href {https://doi.org/10.1021/acscentsci.9b00202} {\bibfield  {journal} {\bibinfo  {journal} {ACS Cent. Sci.}\ }\textbf {\bibinfo {volume} {5}},\ \bibinfo {pages} {900} (\bibinfo {year} {2019})},\ \bibinfo {note} {pMID: 31139726}\BibitemShut {NoStop}%
\bibitem [{\citenamefont {Ma}\ \emph {et~al.}(2020)\citenamefont {Ma}, \citenamefont {Wang}, \citenamefont {Nie}, \citenamefont {Yi}, \citenamefont {Xu}, \citenamefont {Li}, \citenamefont {Jandke}, \citenamefont {Wulfhekel}, \citenamefont {Huang}, \citenamefont {West}, \citenamefont {Richard}, \citenamefont {Chikina}, \citenamefont {Strocov}, \citenamefont {Mesot}, \citenamefont {Weng}, \citenamefont {Zhang}, \citenamefont {Shi}, \citenamefont {Qian}, \citenamefont {Shi},\ and\ \citenamefont {Ding}}]{JunzhangEuCd2As}%
  \BibitemOpen
  \bibfield  {author} {\bibinfo {author} {\bibfnamefont {J.}~\bibnamefont {Ma}}, \bibinfo {author} {\bibfnamefont {H.}~\bibnamefont {Wang}}, \bibinfo {author} {\bibfnamefont {S.}~\bibnamefont {Nie}}, \bibinfo {author} {\bibfnamefont {C.}~\bibnamefont {Yi}}, \bibinfo {author} {\bibfnamefont {Y.}~\bibnamefont {Xu}}, \bibinfo {author} {\bibfnamefont {H.}~\bibnamefont {Li}}, \bibinfo {author} {\bibfnamefont {J.}~\bibnamefont {Jandke}}, \bibinfo {author} {\bibfnamefont {W.}~\bibnamefont {Wulfhekel}}, \bibinfo {author} {\bibfnamefont {Y.}~\bibnamefont {Huang}}, \bibinfo {author} {\bibfnamefont {D.}~\bibnamefont {West}}, \bibinfo {author} {\bibfnamefont {P.}~\bibnamefont {Richard}}, \bibinfo {author} {\bibfnamefont {A.}~\bibnamefont {Chikina}}, \bibinfo {author} {\bibfnamefont {V.~N.}\ \bibnamefont {Strocov}}, \bibinfo {author} {\bibfnamefont {J.}~\bibnamefont {Mesot}}, \bibinfo {author} {\bibfnamefont {H.}~\bibnamefont {Weng}}, \bibinfo {author} {\bibfnamefont {S.}~\bibnamefont {Zhang}}, \bibinfo {author}
  {\bibfnamefont {Y.}~\bibnamefont {Shi}}, \bibinfo {author} {\bibfnamefont {T.}~\bibnamefont {Qian}}, \bibinfo {author} {\bibfnamefont {M.}~\bibnamefont {Shi}},\ and\ \bibinfo {author} {\bibfnamefont {H.}~\bibnamefont {Ding}},\ }\href {https://doi.org/https://doi.org/10.1002/adma.201907565} {\bibfield  {journal} {\bibinfo  {journal} {Adv. Mater.}\ }\textbf {\bibinfo {volume} {32}},\ \bibinfo {pages} {1907565} (\bibinfo {year} {2020})}\BibitemShut {NoStop}%
\bibitem [{\citenamefont {Su}\ \emph {et~al.}(2020)\citenamefont {Su}, \citenamefont {Gong}, \citenamefont {Shi}, \citenamefont {Yang}, \citenamefont {Wang}, \citenamefont {Xia}, \citenamefont {Yu}, \citenamefont {Guo}, \citenamefont {Wang}, \citenamefont {Ding}, \citenamefont {Xu}, \citenamefont {Li}, \citenamefont {Wang}, \citenamefont {Zou}, \citenamefont {Yu}, \citenamefont {Zhu}, \citenamefont {Chen}, \citenamefont {Liu}, \citenamefont {Liu}, \citenamefont {Li},\ and\ \citenamefont {Guo}}]{SuEuCd2Sb2}%
  \BibitemOpen
  \bibfield  {author} {\bibinfo {author} {\bibfnamefont {H.}~\bibnamefont {Su}}, \bibinfo {author} {\bibfnamefont {B.}~\bibnamefont {Gong}}, \bibinfo {author} {\bibfnamefont {W.}~\bibnamefont {Shi}}, \bibinfo {author} {\bibfnamefont {H.}~\bibnamefont {Yang}}, \bibinfo {author} {\bibfnamefont {H.}~\bibnamefont {Wang}}, \bibinfo {author} {\bibfnamefont {W.}~\bibnamefont {Xia}}, \bibinfo {author} {\bibfnamefont {Z.}~\bibnamefont {Yu}}, \bibinfo {author} {\bibfnamefont {P.-J.}\ \bibnamefont {Guo}}, \bibinfo {author} {\bibfnamefont {J.}~\bibnamefont {Wang}}, \bibinfo {author} {\bibfnamefont {L.}~\bibnamefont {Ding}}, \bibinfo {author} {\bibfnamefont {L.}~\bibnamefont {Xu}}, \bibinfo {author} {\bibfnamefont {X.}~\bibnamefont {Li}}, \bibinfo {author} {\bibfnamefont {X.}~\bibnamefont {Wang}}, \bibinfo {author} {\bibfnamefont {Z.}~\bibnamefont {Zou}}, \bibinfo {author} {\bibfnamefont {N.}~\bibnamefont {Yu}}, \bibinfo {author} {\bibfnamefont {Z.}~\bibnamefont {Zhu}}, \bibinfo {author} {\bibfnamefont {Y.}~\bibnamefont
  {Chen}}, \bibinfo {author} {\bibfnamefont {Z.}~\bibnamefont {Liu}}, \bibinfo {author} {\bibfnamefont {K.}~\bibnamefont {Liu}}, \bibinfo {author} {\bibfnamefont {G.}~\bibnamefont {Li}},\ and\ \bibinfo {author} {\bibfnamefont {Y.}~\bibnamefont {Guo}},\ }\href {https://doi.org/10.1063/1.5129467} {\bibfield  {journal} {\bibinfo  {journal} {APL Mater.}\ }\textbf {\bibinfo {volume} {8}},\ \bibinfo {pages} {011109} (\bibinfo {year} {2020})}\BibitemShut {NoStop}%
\bibitem [{\citenamefont {Yang}\ \emph {et~al.}(2024)\citenamefont {Yang}, \citenamefont {Xia}, \citenamefont {Mi}, \citenamefont {Zhang}, \citenamefont {Zhang}, \citenamefont {Wang}, \citenamefont {Chai}, \citenamefont {Zhou}, \citenamefont {Guo},\ and\ \citenamefont {He}}]{Yang2024}%
  \BibitemOpen
  \bibfield  {author} {\bibinfo {author} {\bibfnamefont {K.}~\bibnamefont {Yang}}, \bibinfo {author} {\bibfnamefont {W.}~\bibnamefont {Xia}}, \bibinfo {author} {\bibfnamefont {X.}~\bibnamefont {Mi}}, \bibinfo {author} {\bibfnamefont {Y.}~\bibnamefont {Zhang}}, \bibinfo {author} {\bibfnamefont {L.}~\bibnamefont {Zhang}}, \bibinfo {author} {\bibfnamefont {A.}~\bibnamefont {Wang}}, \bibinfo {author} {\bibfnamefont {Y.}~\bibnamefont {Chai}}, \bibinfo {author} {\bibfnamefont {X.}~\bibnamefont {Zhou}}, \bibinfo {author} {\bibfnamefont {Y.}~\bibnamefont {Guo}},\ and\ \bibinfo {author} {\bibfnamefont {M.}~\bibnamefont {He}},\ }\href {https://doi.org/10.1063/5.0230915} {\bibfield  {journal} {\bibinfo  {journal} {Appl. Phys. Lett.}\ }\textbf {\bibinfo {volume} {125}},\ \bibinfo {pages} {171901} (\bibinfo {year} {2024})}\BibitemShut {NoStop}%
\bibitem [{\citenamefont {Mozaffari}\ \emph {et~al.}(2024)\citenamefont {Mozaffari}, \citenamefont {Meier}, \citenamefont {Madhogaria}, \citenamefont {Peshcherenko}, \citenamefont {Kang}, \citenamefont {Villanova}, \citenamefont {Arachchige}, \citenamefont {Zheng}, \citenamefont {Zhu}, \citenamefont {Chen}, \citenamefont {Jenkins}, \citenamefont {Zhang}, \citenamefont {Chan}, \citenamefont {Li}, \citenamefont {Yoon}, \citenamefont {Zhang},\ and\ \citenamefont {Mandrus}}]{Mo_Sc}%
  \BibitemOpen
  \bibfield  {author} {\bibinfo {author} {\bibfnamefont {S.}~\bibnamefont {Mozaffari}}, \bibinfo {author} {\bibfnamefont {W.~R.}\ \bibnamefont {Meier}}, \bibinfo {author} {\bibfnamefont {R.~P.}\ \bibnamefont {Madhogaria}}, \bibinfo {author} {\bibfnamefont {N.}~\bibnamefont {Peshcherenko}}, \bibinfo {author} {\bibfnamefont {S.-H.}\ \bibnamefont {Kang}}, \bibinfo {author} {\bibfnamefont {J.~W.}\ \bibnamefont {Villanova}}, \bibinfo {author} {\bibfnamefont {H.~W.~S.}\ \bibnamefont {Arachchige}}, \bibinfo {author} {\bibfnamefont {G.}~\bibnamefont {Zheng}}, \bibinfo {author} {\bibfnamefont {Y.}~\bibnamefont {Zhu}}, \bibinfo {author} {\bibfnamefont {K.-W.}\ \bibnamefont {Chen}}, \bibinfo {author} {\bibfnamefont {K.}~\bibnamefont {Jenkins}}, \bibinfo {author} {\bibfnamefont {D.}~\bibnamefont {Zhang}}, \bibinfo {author} {\bibfnamefont {A.}~\bibnamefont {Chan}}, \bibinfo {author} {\bibfnamefont {L.}~\bibnamefont {Li}}, \bibinfo {author} {\bibfnamefont {M.}~\bibnamefont {Yoon}}, \bibinfo {author} {\bibfnamefont
  {Y.}~\bibnamefont {Zhang}},\ and\ \bibinfo {author} {\bibfnamefont {D.~G.}\ \bibnamefont {Mandrus}},\ }\href {https://doi.org/10.1103/PhysRevB.110.035135} {\bibfield  {journal} {\bibinfo  {journal} {Phys. Rev. B}\ }\textbf {\bibinfo {volume} {110}},\ \bibinfo {pages} {035135} (\bibinfo {year} {2024})}\BibitemShut {NoStop}%
\bibitem [{\citenamefont {DeStefano}\ \emph {et~al.}(2023)\citenamefont {DeStefano}, \citenamefont {Rosenberg}, \citenamefont {Peek}, \citenamefont {Lee}, \citenamefont {Liu}, \citenamefont {Jiang}, \citenamefont {Ke},\ and\ \citenamefont {Chu}}]{DeStefano_Sc}%
  \BibitemOpen
  \bibfield  {author} {\bibinfo {author} {\bibfnamefont {J.~M.}\ \bibnamefont {DeStefano}}, \bibinfo {author} {\bibfnamefont {E.}~\bibnamefont {Rosenberg}}, \bibinfo {author} {\bibfnamefont {O.}~\bibnamefont {Peek}}, \bibinfo {author} {\bibfnamefont {Y.}~\bibnamefont {Lee}}, \bibinfo {author} {\bibfnamefont {Z.}~\bibnamefont {Liu}}, \bibinfo {author} {\bibfnamefont {Q.}~\bibnamefont {Jiang}}, \bibinfo {author} {\bibfnamefont {L.}~\bibnamefont {Ke}},\ and\ \bibinfo {author} {\bibfnamefont {J.-H.}\ \bibnamefont {Chu}},\ }\href {https://doi.org/10.1038/s41535-023-00600-8} {\bibfield  {journal} {\bibinfo  {journal} {npj Quantum Materials}\ }\textbf {\bibinfo {volume} {8}},\ \bibinfo {pages} {65} (\bibinfo {year} {2023})}\BibitemShut {NoStop}%
\bibitem [{\citenamefont {Chen}\ \emph {et~al.}(2023)\citenamefont {Chen}, \citenamefont {Liu}, \citenamefont {Xia}, \citenamefont {Mi}, \citenamefont {Zhong}, \citenamefont {Yang}, \citenamefont {Zhang}, \citenamefont {Gan}, \citenamefont {Liu}, \citenamefont {Wang}, \citenamefont {Wang}, \citenamefont {Chai}, \citenamefont {Shen}, \citenamefont {Yang}, \citenamefont {Guo},\ and\ \citenamefont {He}}]{Chen_Ti3Bi5}%
  \BibitemOpen
  \bibfield  {author} {\bibinfo {author} {\bibfnamefont {X.}~\bibnamefont {Chen}}, \bibinfo {author} {\bibfnamefont {X.}~\bibnamefont {Liu}}, \bibinfo {author} {\bibfnamefont {W.}~\bibnamefont {Xia}}, \bibinfo {author} {\bibfnamefont {X.}~\bibnamefont {Mi}}, \bibinfo {author} {\bibfnamefont {L.}~\bibnamefont {Zhong}}, \bibinfo {author} {\bibfnamefont {K.}~\bibnamefont {Yang}}, \bibinfo {author} {\bibfnamefont {L.}~\bibnamefont {Zhang}}, \bibinfo {author} {\bibfnamefont {Y.}~\bibnamefont {Gan}}, \bibinfo {author} {\bibfnamefont {Y.}~\bibnamefont {Liu}}, \bibinfo {author} {\bibfnamefont {G.}~\bibnamefont {Wang}}, \bibinfo {author} {\bibfnamefont {A.}~\bibnamefont {Wang}}, \bibinfo {author} {\bibfnamefont {Y.}~\bibnamefont {Chai}}, \bibinfo {author} {\bibfnamefont {J.}~\bibnamefont {Shen}}, \bibinfo {author} {\bibfnamefont {X.}~\bibnamefont {Yang}}, \bibinfo {author} {\bibfnamefont {Y.}~\bibnamefont {Guo}},\ and\ \bibinfo {author} {\bibfnamefont {M.}~\bibnamefont {He}},\ }\href
  {https://doi.org/10.1103/PhysRevB.107.174510} {\bibfield  {journal} {\bibinfo  {journal} {Phys. Rev. B}\ }\textbf {\bibinfo {volume} {107}},\ \bibinfo {pages} {174510} (\bibinfo {year} {2023})}\BibitemShut {NoStop}%
\bibitem [{\citenamefont {Kohler}(1938)}]{Kohler}%
  \BibitemOpen
  \bibfield  {author} {\bibinfo {author} {\bibfnamefont {M.}~\bibnamefont {Kohler}},\ }\href {https://doi.org/https://doi.org/10.1002/andp.19384240124} {\bibfield  {journal} {\bibinfo  {journal} {Annalen der Physik}\ }\textbf {\bibinfo {volume} {424}},\ \bibinfo {pages} {211} (\bibinfo {year} {1938})}\BibitemShut {NoStop}%
\bibitem [{\citenamefont {Harris}\ \emph {et~al.}(1995)\citenamefont {Harris}, \citenamefont {Yan}, \citenamefont {Matl}, \citenamefont {Ong}, \citenamefont {Anderson}, \citenamefont {Kimura},\ and\ \citenamefont {Kitazawa}}]{Harris_vio}%
  \BibitemOpen
  \bibfield  {author} {\bibinfo {author} {\bibfnamefont {J.~M.}\ \bibnamefont {Harris}}, \bibinfo {author} {\bibfnamefont {Y.~F.}\ \bibnamefont {Yan}}, \bibinfo {author} {\bibfnamefont {P.}~\bibnamefont {Matl}}, \bibinfo {author} {\bibfnamefont {N.~P.}\ \bibnamefont {Ong}}, \bibinfo {author} {\bibfnamefont {P.~W.}\ \bibnamefont {Anderson}}, \bibinfo {author} {\bibfnamefont {T.}~\bibnamefont {Kimura}},\ and\ \bibinfo {author} {\bibfnamefont {K.}~\bibnamefont {Kitazawa}},\ }\href {https://doi.org/10.1103/PhysRevLett.75.1391} {\bibfield  {journal} {\bibinfo  {journal} {Phys. Rev. Lett.}\ }\textbf {\bibinfo {volume} {75}},\ \bibinfo {pages} {1391} (\bibinfo {year} {1995})}\BibitemShut {NoStop}%
\bibitem [{\citenamefont {Anderson}(1997)}]{Anderson_vio}%
  \BibitemOpen
  \bibfield  {author} {\bibinfo {author} {\bibfnamefont {P.~W.}\ \bibnamefont {Anderson}},\ }\href {https://doi.org/10.1063/1.881959} {\bibfield  {journal} {\bibinfo  {journal} {Phys. Today}\ }\textbf {\bibinfo {volume} {50}},\ \bibinfo {pages} {42} (\bibinfo {year} {1997})}\BibitemShut {NoStop}%
\bibitem [{\citenamefont {Li}\ \emph {et~al.}(2018)\citenamefont {Li}, \citenamefont {Sun}, \citenamefont {Shahi}, \citenamefont {Gao}, \citenamefont {MacDonald}, \citenamefont {Uwatoko}, \citenamefont {Xiang}, \citenamefont {Goodenough}, \citenamefont {Cheng},\ and\ \citenamefont {Zhou}}]{Xiang_vio}%
  \BibitemOpen
  \bibfield  {author} {\bibinfo {author} {\bibfnamefont {X.}~\bibnamefont {Li}}, \bibinfo {author} {\bibfnamefont {J.}~\bibnamefont {Sun}}, \bibinfo {author} {\bibfnamefont {P.}~\bibnamefont {Shahi}}, \bibinfo {author} {\bibfnamefont {M.}~\bibnamefont {Gao}}, \bibinfo {author} {\bibfnamefont {A.~H.}\ \bibnamefont {MacDonald}}, \bibinfo {author} {\bibfnamefont {Y.}~\bibnamefont {Uwatoko}}, \bibinfo {author} {\bibfnamefont {T.}~\bibnamefont {Xiang}}, \bibinfo {author} {\bibfnamefont {J.~B.}\ \bibnamefont {Goodenough}}, \bibinfo {author} {\bibfnamefont {J.}~\bibnamefont {Cheng}},\ and\ \bibinfo {author} {\bibfnamefont {J.}~\bibnamefont {Zhou}},\ }\href {https://doi.org/10.1073/pnas.1810726115} {\bibfield  {journal} {\bibinfo  {journal} {Proc. Natl. Acad. Sci. U.S.A.}\ }\textbf {\bibinfo {volume} {115}},\ \bibinfo {pages} {9935} (\bibinfo {year} {2018})}\BibitemShut {NoStop}%
\bibitem [{\citenamefont {R\"o\ss{}ler}\ \emph {et~al.}(2015)\citenamefont {R\"o\ss{}ler}, \citenamefont {Koz}, \citenamefont {Jiao}, \citenamefont {R\"o\ss{}ler}, \citenamefont {Steglich}, \citenamefont {Schwarz},\ and\ \citenamefont {Wirth}}]{R_vio}%
  \BibitemOpen
  \bibfield  {author} {\bibinfo {author} {\bibfnamefont {S.}~\bibnamefont {R\"o\ss{}ler}}, \bibinfo {author} {\bibfnamefont {C.}~\bibnamefont {Koz}}, \bibinfo {author} {\bibfnamefont {L.}~\bibnamefont {Jiao}}, \bibinfo {author} {\bibfnamefont {U.~K.}\ \bibnamefont {R\"o\ss{}ler}}, \bibinfo {author} {\bibfnamefont {F.}~\bibnamefont {Steglich}}, \bibinfo {author} {\bibfnamefont {U.}~\bibnamefont {Schwarz}},\ and\ \bibinfo {author} {\bibfnamefont {S.}~\bibnamefont {Wirth}},\ }\href {https://doi.org/10.1103/PhysRevB.92.060505} {\bibfield  {journal} {\bibinfo  {journal} {Phys. Rev. B}\ }\textbf {\bibinfo {volume} {92}},\ \bibinfo {pages} {060505} (\bibinfo {year} {2015})}\BibitemShut {NoStop}%
\bibitem [{\citenamefont {Wu}\ \emph {et~al.}(2015)\citenamefont {Wu}, \citenamefont {Jo}, \citenamefont {Ochi}, \citenamefont {Huang}, \citenamefont {Mou}, \citenamefont {Bud'ko}, \citenamefont {Canfield}, \citenamefont {Trivedi}, \citenamefont {Arita},\ and\ \citenamefont {Kaminski}}]{Wu_vio}%
  \BibitemOpen
  \bibfield  {author} {\bibinfo {author} {\bibfnamefont {Y.}~\bibnamefont {Wu}}, \bibinfo {author} {\bibfnamefont {N.~H.}\ \bibnamefont {Jo}}, \bibinfo {author} {\bibfnamefont {M.}~\bibnamefont {Ochi}}, \bibinfo {author} {\bibfnamefont {L.}~\bibnamefont {Huang}}, \bibinfo {author} {\bibfnamefont {D.}~\bibnamefont {Mou}}, \bibinfo {author} {\bibfnamefont {S.~L.}\ \bibnamefont {Bud'ko}}, \bibinfo {author} {\bibfnamefont {P.~C.}\ \bibnamefont {Canfield}}, \bibinfo {author} {\bibfnamefont {N.}~\bibnamefont {Trivedi}}, \bibinfo {author} {\bibfnamefont {R.}~\bibnamefont {Arita}},\ and\ \bibinfo {author} {\bibfnamefont {A.}~\bibnamefont {Kaminski}},\ }\href {https://doi.org/10.1103/PhysRevLett.115.166602} {\bibfield  {journal} {\bibinfo  {journal} {Phys. Rev. Lett.}\ }\textbf {\bibinfo {volume} {115}},\ \bibinfo {pages} {166602} (\bibinfo {year} {2015})}\BibitemShut {NoStop}%
\bibitem [{\citenamefont {Xu}\ \emph {et~al.}(2021{\natexlab{b}})\citenamefont {Xu}, \citenamefont {Han}, \citenamefont {Wang}, \citenamefont {Thoutam}, \citenamefont {Pate}, \citenamefont {Li}, \citenamefont {Zhang}, \citenamefont {Wang}, \citenamefont {Fotovat}, \citenamefont {Welp}, \citenamefont {Zhou}, \citenamefont {Kwok}, \citenamefont {Chung}, \citenamefont {Kanatzidis},\ and\ \citenamefont {Xiao}}]{Xu_vio}%
  \BibitemOpen
  \bibfield  {author} {\bibinfo {author} {\bibfnamefont {J.}~\bibnamefont {Xu}}, \bibinfo {author} {\bibfnamefont {F.}~\bibnamefont {Han}}, \bibinfo {author} {\bibfnamefont {T.-T.}\ \bibnamefont {Wang}}, \bibinfo {author} {\bibfnamefont {L.~R.}\ \bibnamefont {Thoutam}}, \bibinfo {author} {\bibfnamefont {S.~E.}\ \bibnamefont {Pate}}, \bibinfo {author} {\bibfnamefont {M.}~\bibnamefont {Li}}, \bibinfo {author} {\bibfnamefont {X.}~\bibnamefont {Zhang}}, \bibinfo {author} {\bibfnamefont {Y.-L.}\ \bibnamefont {Wang}}, \bibinfo {author} {\bibfnamefont {R.}~\bibnamefont {Fotovat}}, \bibinfo {author} {\bibfnamefont {U.}~\bibnamefont {Welp}}, \bibinfo {author} {\bibfnamefont {X.}~\bibnamefont {Zhou}}, \bibinfo {author} {\bibfnamefont {W.-K.}\ \bibnamefont {Kwok}}, \bibinfo {author} {\bibfnamefont {D.~Y.}\ \bibnamefont {Chung}}, \bibinfo {author} {\bibfnamefont {M.~G.}\ \bibnamefont {Kanatzidis}},\ and\ \bibinfo {author} {\bibfnamefont {Z.-L.}\ \bibnamefont {Xiao}},\ }\href {https://doi.org/10.1103/PhysRevX.11.041029}
  {\bibfield  {journal} {\bibinfo  {journal} {Phys. Rev. X}\ }\textbf {\bibinfo {volume} {11}},\ \bibinfo {pages} {041029} (\bibinfo {year} {2021}{\natexlab{b}})}\BibitemShut {NoStop}%
\end{thebibliography}

\end{document}